\documentclass[floatfix, prl, twocolumn, showpacs, showkeys, preprintnumbers, nofootinbib, superscriptaddress]{revtex4-1}
\usepackage[utf8]{inputenc}
\usepackage[sort&compress]{natbib}
\usepackage{ulem}
\usepackage{bm}
\usepackage{times}
\usepackage{amssymb,amsbsy,amsmath,amsfonts}
\usepackage{graphicx}
\usepackage{float}
\usepackage{color}
\usepackage{morefloats}
\usepackage{rotating}
\usepackage{srcltx}
\usepackage{slashed}
\usepackage{subfigure}
\usepackage{multirow}
\usepackage{verbatim}
\usepackage{hyperref}
\usepackage{tabularx}
\usepackage{braket}

\DeclareUnicodeCharacter{3000}{HEREHEREHERE}

\begin{document}

\title{Theoretical investigation of the molecular nature of $D_{s0}^*(2317)$ and $D_{s1}(2460)$ and the possibility of observing  the $D\bar{D}K$ bound state $K_{c\bar{c}}(4180)$ in inclusive $e^+e^-\to c\bar{c}$ collisions}

\author{Tian-Chen Wu}
\affiliation{School of Physics, Beihang University, Beijing, 102206, China}

\author{Li-Sheng Geng}
\email[Corresponding author: ]{lisheng.geng@buaa.edu.cn}

\affiliation{School of Physics, Beihang University, Beijing, 102206, China}
\affiliation{Peng Huanwu Collaborative Center for Research and Education, Beihang University, Beijing 100191, China}
\affiliation{Beijing Key Laboratory of Advanced Nuclear Materials and Physics, Beihang University, Beijing, 102206, China}
\affiliation{School of Physics and Microelectronics, Zhengzhou University, Zhengzhou, Henan, 450001, China}

\begin{abstract}
Searching for exotic multiquark states and elucidating their nature remains a central topic in   
understanding quantum chromodynamics--the underlying theory of the strong interaction. Two of the most studied such states are  the charm-strange states $D_{s0}^*(2317)$ and $D_{s1}(2460)$.  In this letter,  we show for the first time that   their prompt production yields  in inclusive $e^+e^-\to c\bar{c}$ collisions near $\sqrt{s}=10.6$ GeV measured by the BABAR Collaboration,  $Y(D_{s0}^*(2317))$ and $Y(D_{s1}(2460))$, in particular the ratio $R=Y(D_{s0}^*(2317))/Y(D_{s1}(2460))$, can be well explained in the molecular picture, which provide a highly nontrivial verification of their nature being $DK/D^*K$ molecules. On the contrary, treating them as pure $c\bar{s}$ $P-$wave states, the statistical model predicts a ratio $R$ smaller than unity, in contrast with the experimental central value, though in agreement with it considering its relatively large uncertainty.  In addition, we predict the production yield of  the  $D\bar{D}K$ three-body bound state, $K_{c\bar{c}}(4180)$, in $e^+e^-\to c\bar{c}$ collisions and find that it is within the reach of the ongoing Belle II experiment. The present study demonstrates the feasibility of a novel method to unravel the nature of exotic hadrons and  the potential of electron-positron collisions in this regard. 
\end{abstract}

\maketitle

{\noindent \it Introduction --} In 2003,  two exotic hadrons  were discovered, i.e., $X(3872)$~\cite{Belle:2003nnu} and $D_{s0}^*(2317)$~\cite{BaBar:2003oey2317},  which cannot be  easily accommodated in the conventional quark model  of Gell-Mann and Zweig~\cite{Gell-Mann:1964ewy,Zweig:1964jf} where  baryons  and mesons are viewed as three quark and quark-antiquark color singlets, respectively.  
In the following years, more such hadrons were discovered~\cite{Guo:2017jvc,Olsen:2017bmm,Ali:2017jda,Brambilla:2019esw,Chen:2021ftn,Chen:2022asf}, some of which, because of their proximity to the mass thresholds of pairs of conventional hadrons, are often interpreted as hadronic molecules~\cite{Guo:2017jvc}, i.e., multi-hadron states bound by the residual strong nuclear force instead of the electromagnetic interaction. In particular, the $D_{s0}^*(2317)$~\cite{BaBar:2003oey2317,CLEO:2003ggt2317,Belle:2003guh2317} and its heavy-spin symmetry partner $D_{s1}(2460)$~\cite{CLEO:2003ggt2317,Belle:2003guh2317}, with masses lower by 160 MeV and 70 MeV  than their counterparts in the  Godfrey-Isgur (GI) quark model~\cite{Godfrey:1985xj}, are shown to qualify as bound states of $DK$ and $D^*K$~\cite{Kolomeitsev:2003molecule,vanBeveren:2003molecule,Barnes:2003molecule,Chen:2004molecule,Guo:2006molecule2,Liu:2012lattice,Mohler:2013lattice,Lang:2014lattice,Altenbuchinger:2013vwa,MartinezTorres:2014kpc,Yang:2021tvc,Faessler:2007cu,Liu:2022zbd}. It should be mentioned although the molecular picture for  $D_{s0}^*(2317)/D_{s1}(2460)$ is the prevailing  one, alternative interpretations do exist, such as conventional $q\bar{q}$ states~\cite{Dai:2003yg,Lakhina:2006fy} or compact tetraquark states~\cite{Barnes:2003dj} or mixtures of them~\cite{Browder:2003fk}.

By analogy with the only well-established “hardronic molecules” in nature, i.e., atomic nuclei and hypernuclei, it was realized recently that one can build heavier three-body even four-body systems starting from the fact that the $DK$/$D^*K$ interactions are attractive and strong enough to form the $D_{s0}^*(2317)$ and $D_{s1}(2460)$ states~\cite{Wu:2019vsy,MartinezTorres:2018zbl,Ren:2018pcd}. Note that this picture has attracted considerable attention and many similar studies have been performed and several good candidates have been identified~\cite{Wu:2021dwy,MartinezTorres:2020hus,Wu:2022ftm,Wu:2021gyn,Wu:2020job,Wu:2020rdg,Wu:2021kbu,Luo:2021ggs,Pan:2022xxz}.  The next step forward is to search for their existence experimentally. Recently, the Belle Collaboration has performed the first dedicated search for the predicted $DDK$ bound state in $e^+e^-$ collisions~\cite{Belle:2020xca}. Further theoretical studies are in urgent need to verify the above conjecture~\cite{Wu:2019vsy,MartinezTorres:2018zbl,Ren:2018pcd,Wu:2021dwy,MartinezTorres:2020hus,Wu:2022ftm,Wu:2021gyn,Wu:2020job,Wu:2020rdg,Wu:2021kbu,Luo:2021ggs,Pan:2022xxz} and guide future experiments. 

The nucleus(molecule)-like nature of the $D_{s0}^*(2317)$ and $D_{s1}(2460)$ suggests that one should be able to explain their prompt production yields using the coalescence mechanism, which is known to work well in reproducing the productions of light (anti)nuclei and hyper nuclei in heavy-ion and hadron-hadron collisions~\cite{Mattiello:1996wignertheory3,Nagle:1996cutoff4Wigner,STAR:2011eej,ALICE:2022jmr,STAR:2010gyg,Gev:2022ksw}. In the coalescence mechanism,  composite particles are formed by coalescence of their constituents that satisfy the constraints in phase space at kinetic freeze-out. From such a perspective, the production yields of exotic hadrons in hadron-hadron and heavy-ion collisions have been studied~\cite{Exotic:2010relativistic5coalescence,Zhang:2020dwn,Chen:PACIAE_3872_2022,Chen:2021akx,Hu:2021gdg, Wu:2020zbx, Abreu:2022lfy,Yoon:2022voo},  but similar studies in electron-positron collisions are much rare. In particular, 
a dedicated study of the yields of $D_{s0}^*(2317)$ and $D_{s1}(2460)$ in $e^+e^-$ collisions, where they were first observed, is still missing. In this work, we fill this gap and study the prompt productions of $D_{s0}^*(2317)$ and $D_{s1}(2460)$ in $e^+e^-$ collisions.  We find that the results are in nice agreement with the BABAR data~\cite{BaBar:2006eep} and therefore provide a highly nontrivial support for their molecular nature. Built on this success, we further predict the production yield of the three-body $D\bar{D}K$ bound state, which is formed by the same $DK$($\bar{D}K$) interaction that binds $D_{s0}^*(2317)$, and find that it is within the reach of the on-going Belle II experiment, which, if found in the future, will not only help further confirm the molecular nature of $D_{s0}^*(2317)$ and $D_{s1}(2460)$ but also open a new chapter in studies of the non-perturbative strong interaction.

{\it Theoretical framework:} 
We adopt the time-honored coalescence model~\cite{Sato:1981ez} to study the production of $DK/D^*K/D\bar{D}K$ molecules. There are two essential ingredients in such a study, i.e., the production of primary hadrons and the coalescence process. For the former, a transport model is used to provide the phase space information of the particle source containing the primary hadrons ($D$, $D^*$, $\bar{D}$, and $K$)  of our interest at the kinetic freeze-out.
For the latter, the Wigner function method is adopted. 

For the transport process, we adopt the PACIAE model~\cite{Sa:2011PACIAE20,Yan:PACIAE_light_nuclei,Chen:PACIAE_3872_2022}, which is a transport model based on the event generator PYTHIA~\cite{Sjostrand:2006PYTHIA}. 
Once the primary hadrons are produced, we apply the coalescence model to study the formation of hadronic molecules. The coalescence model is widely used to calculate the production rates of composite particles such as nuclear clusters and hadronic molecules~\cite{Mattiello:1996wignertheory3,Nagle:1996cutoff4Wigner,Chen:2003deuteronwigner,Sombun:2018yqh,Deng:2020zxo,Zhang:2020diabaryon,Exotic:2010relativistic5coalescence}. The basic idea is that the constituents of a shallow bound composite particle, whose binding energy is small compared to the evolution temperature, only combine together until the whole system reaches the kinetic freeze-out.

The formation of clusters can be described in the final-state interaction approximation~\cite{Gyulassy:1982wignertheory1} (see the Supplemental Material for details), indicating that it  only occurs when the interactions between their constituents almost cease and the formation time  is short compared to their interaction time. Then the production yield is the overlap integral between the cluster density $\hat{\rho}_C$ and the final-state source density $\hat{\rho}_S$.
Both densities need to be transformed into the Wigner densities $\hat{\rho}_S^W$ and $\hat{\rho}_C^W$ since the source density obtained from the transport model is semi-classical. The source Wigner density $\hat{\rho}_S^W$ can be constructed from the positions $\Tilde{\bm x}_{n}$ and momenta $\Tilde{\bm p}_{n}$ of the primary particles after the kinetic freeze-out~\cite{Nagle:1996cutoff4Wigner}, to which each particle $n$ contributes a product of delta functions $\delta^3(\bm{x}_n-\bm{\Tilde{x}}_n)\delta^3(\bm{p}_n-\bm{\Tilde{p}}_n)$. 
The cluster Wigner density $\hat{\rho}_C^W$ can be obtained by the following Wigner transformation
\begin{widetext}
\begin{equation}
\begin{aligned}\label{Eq:cluster density}
\hat{\rho}_C^W(\bm r_1,\bm q_1,\cdots,\bm r_{n-1}, \bm q_{n-1})=&\int\Psi_C(\bm{r}_1+\frac{1}{2}\bm{y}_1,\cdots,\bm{r}_{n-1}+\frac{1}{2}\bm{y}_{n-1})\Psi_C^{\ast}(\bm{r}_1-\frac{1}{2}\bm{y}_1,\cdots,\bm{r}_{n-1}-\frac{1}{2}\bm{y}_{n-1})\\&
 e^{-i\bm{q}_1 \cdot \bm{y}_1}\cdots e^{-i\bm{q}_{n-1} \cdot \bm{y}_{n-1}} d^3\bm y_1\cdots  d^3\bm y_{n-1},
\end{aligned}
\end{equation}
\end{widetext}
where $\Psi_C$ is the relative wave function of the $n$-body cluster defined in its center-of-mass (c.m.) system , $\bm r_{n-1}$ and $\bm q_{n-1}$ are the $n-1$ relative coordinates calculated from space and momentum coordinates $\bm x_{n}$ and $\bm p_{n}$.
Then one obtains the yield of the $n$-body cluster~\cite{Mattiello:1996wignertheory3,Nagle:1996cutoff4Wigner}
\begin{widetext}
\begin{equation}
    \begin{aligned}\label{yield}
 &\frac{dN}{d \bm P}= g \int \hat{\rho}_S^W(\bm x_1,\bm p_1,\cdots,\bm x_n, \bm p_n) \hat{\rho}_C^W(\bm r_1,\bm q_1,\cdots,\bm r_{n-1}, \bm q_{n-1}) \delta^3(\bm P-(\bm p_1+\cdots \bm p_n))\frac{d\bm x_{1} d\bm p_{1} }{(2\pi)^3 } \cdots \frac{d\bm x_{n} d\bm p_{n} }{(2\pi)^3 }  \\
& N= \left \langle \sum_{(\rm c)} \hat{\rho}_C^W(\Tilde{\bm r}_1,\Tilde{\bm q}_1,\cdots,\Tilde{\bm r}_{n-1}, \Tilde{\bm q}_{n-1}) \right \rangle
    \end{aligned}
\end{equation}
\end{widetext}
where $\bm P$ is the total momentum of the cluster, $\Tilde{\bm r}_n$ and $\Tilde{\bm q}_n$ are the relative space  and  momentum coordinates calculated with the space coordinate $\tilde{\bm x}_n$ and momentum coordinate $\tilde{\bm p}_n$ of the primary hadrons in each combination (c) from the transport model, and $\left\langle\cdots\right\rangle$ denotes that the result is averaged over all the event runs. We note that the spin statistical factor $g$ for $DK$, $D^*K$,  and $D\bar{D}K$ is 1.

To compute the cluster Wigner density $\hat{\rho}_C^W$, we need the wave functions of $D_{s0}^*(2317)$ and $D_{s1}(2460)$ as $DK$ and $D^*K$ bound states, which have been studied thoroughly in the chiral unitary approaches at leading order~\cite{Kolomeitsev:2003ac, Gamermann:2006nm, Guo:2006fu}, next-to leading order ~\cite{Altenbuchinger:2013vwa, Liu:2012zya, Guo:2015dha} and next-to next-to leading order~\cite{Yao:2015qia,Du:2017ttu,Huang:2022cag}. Here we follow Refs.~\cite{Wu:2019vsy,Wu:2021dwy} and adopt the wave functions  in coordinate space obtained in the Gaussian expansion method.
 The radial part of the relative wave function can be expanded as a sum of Gaussian  functions,
\begin{equation}
\begin{aligned}\label{Eq:gaussian}
 \Psi_{D^{(*)}K}(\bm{r})= \sum_{i=1}^N c_i(\frac{2\omega_i}{\pi})^{3/4} e^{-\omega_i r^2}, 
 \end{aligned}
 \end{equation}
 where  $\omega_i$ is the width parameter of each basis, $c_i$ is the coefficient, and $N$ is the number of Gaussian  functions. 
With these wave functions, one can easily obtain the Wigner densities and yields for the $D^{(*)}K$ bound states. We stress that the binding energies of $D_{s0}^*(2317)$ and $D_{s1}(2460)$ as $DK$ and $D^*K$ bound states alone cannot uniquely determine their wave functions. Therefore, following Refs.~\cite{Wu:2019vsy,Wu:2021dwy}, we employ three representative wave functions from differently regularized potentials (see the Supplemental Material for more details)  and study the corresponding impact on the production yields.

\begin{figure}[htpb]
\centering
\includegraphics[width=0.48\textwidth]{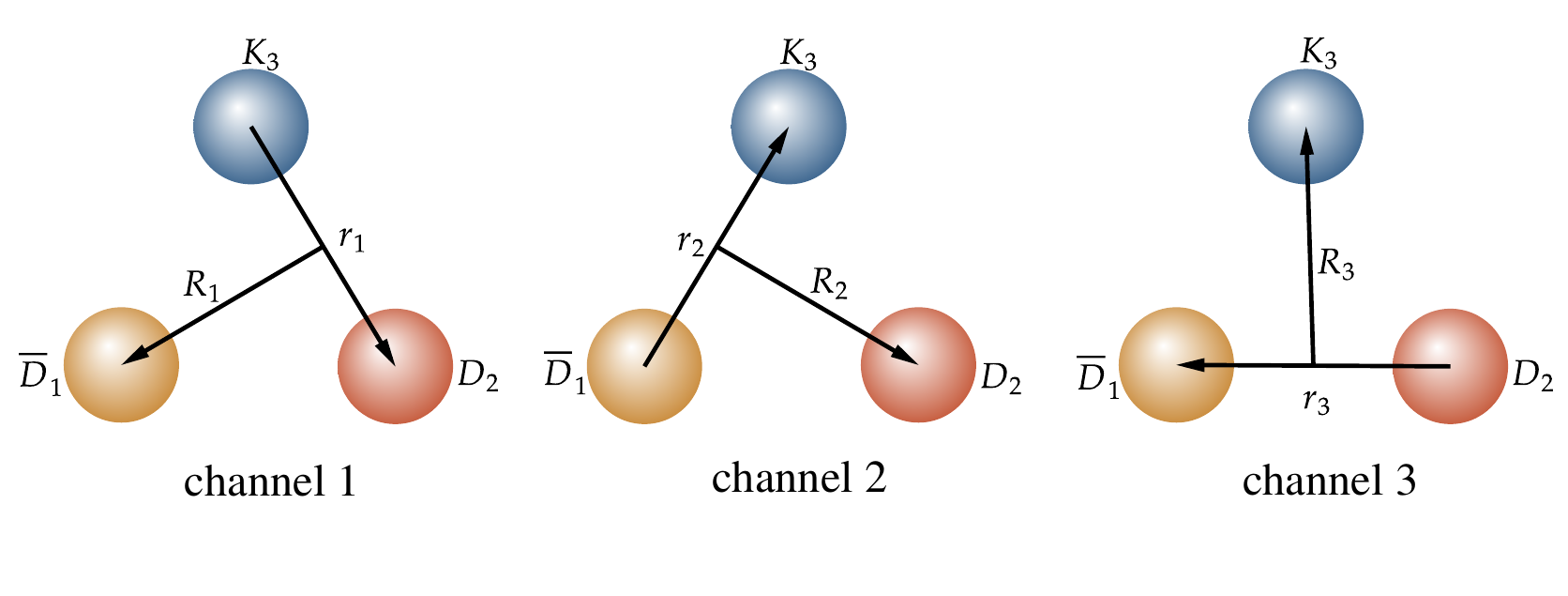}
\caption{ \label{fig:Jacobi}  Jacobi coordinates of the $D\bar{D}K$ system}
\end{figure}
\begin{table}[htpb]
\caption{\label{tab:table5}Experimental and simulated yields (per event) of primary hadrons in $e^+e^-$ annihilations. 
$D$ mesons are measured near $\sqrt{s}$ = 10.5 GeV and simulated at $\sqrt{s}$ = 10.52 GeV in the $e^+e^- \to c\bar{c}$ process.}
\setlength{\tabcolsep}{3.2pt}
\begin{tabular}{ccc}
\hline
\hline
 Particle&Data\cite{Lisovyi:2015uqa}&PACIAE results\\ \hline
 $D^+$ &0.2639$\pm$0.0139 &0.2386\\
 $D^0$ &0.5772$\pm$0.0241 &0.5276\\
$D^{*+}$ &0.2470$\pm$0.0137 &0.2100\\
$D^{*0}$ &0.2241$\pm$0.0304 &0.2026\\
\hline
\hline
\end{tabular}
\end{table}

Following the same technique, one can calculate the production yield of the $D\bar{D}K$ state predicted in Ref.~\cite{Wu:2021dwy}, later confirmed in Ref.~\cite{Wei:2022jgc}. In Ref.~\cite{Wu:2021dwy}, the relative wave function of $D\bar{D}K$ is written in terms of three Jacobi coordinates, as shown in Fig.~\ref{fig:Jacobi}, which reads~\cite{Wu:2021dwy}
 \begin{equation}
 \begin{aligned}\label{Eq:gaussian}
 \Psi_{D\bar{D}K}=& \sum_{\rm ch_i=1}^3  \sum_{i,j=1}^N  c_{i,j}^{(\rm ch_i)} (\frac{4\omega_i\omega_j}{\pi^2})^{3/4}\\
 &\exp(-\omega_i \bm r_{\rm ch_i}^2) \exp(-\omega_j \bm R_{\rm ch_i}^2).
 \end{aligned}
 \end{equation}
 where $\rm ch_i$ is  the label of the three Jacobi channels shown in
 Fig.~\ref{fig:Jacobi}. Using this wave function,  with a bit of algebra, one can obtain the corresponding Wigner density and yield.

\begin{table*}[htpb]
\caption{\label{tab:table6}Yields (per $e^+e^-\to c\bar{c}$ event, containing charge conjugated states) of $D_{s0}(2317)$ and $D_{s1}(2460)$ in $e^+e^-$ collisions and their ratio at $\sqrt{s}$ = 10.58 GeV obtained in the Wigner function approach for c.m. momentum $p^* > 3.2$ GeV/c. The experimental data are estimated from the BABAR data in inclusive $c\bar{c}$ productions near 10.6 GeV~\cite{BaBar:2006eep}. The uncertainty of the ratio is the standard deviation of the three ratios calculated from the central and two boundary values of $\rm r_p$.}
\setlength{\tabcolsep}{3.2pt}
\begin{tabular}{cccccc}
\hline
\hline
&Case 1& Case 2&Case 3&Data~\cite{BaBar:2006eep}\\ \hline
Y$_{D_{s0}(2317)}$& $5.87_{-2.27}^{+3.54}\times10^{-3}$&$4.72_{-1.44}^{+1.60}\times10^{-3}$ &$3.43_{-0.84}^{+0.75}\times10^{-3} $&$5.37_{-1.75}^{+1.39}\times10^{-3}$\\
Y$_{D_{s1}(2460)}$&$3.68_{-1.32}^{+2.43}\times10^{-3}$&$2.98_{-0.73}^{+1.10}\times10^{-3}$&$2.15_{-0.33}^{+0.56}\times10^{-3}$&$3.91_{-1.43}^{+1.43}\times10^{-3}$\\
Y$_{D_{s0}(2317)}$/Y$_{D_{s1}(2460)}$&$1.56 \pm 0.04$&$1.53 \pm 0.07$&$1.53 \pm 0.09$&$1.37_{-0.47}^{+0.39}$~\footnote{The uncertainty of the ratio is calculated from the cross sections directly, not from the yields.}\\
\hline
\hline
\end{tabular}
\end{table*}

It is  also instructive to study the production yields of $D_{s0}^*(2317)$ and $D_{s1}(2460)$ assuming that they are conventional $c\bar{s}$ $P$-wave states. This can be done in the statistical model~\cite{Becattini:1995if}. The details of the statistical model can be found in Ref.~\cite{Becattini:1995if} and those relevant to the present work are given in the Supplemental Material.

{\it Results and Discussion:} First, we estimate the production yields of  $D_{s0}^*(2317)$ and $D_{s1}(2460)$ treated as hadronic molecules of $DK$ and $D^*K$ in our transport plus coalescence model.  Since $D_{s0}^*(2317)$ and $D_{s1}(2460)$ were observed in inclusive $e^+e^-\to c\bar{c}$ collisions at a c.m. energy around $\Upsilon(4S)$~\cite{BaBar:2006eep}, we use the $e^+e^-\to c\bar{c}$ mode in PACIAE to  simulate this process. All the parameters in this mode are fixed at their default values, except for parj(13), the probability that a charm or heavier meson has spin 1. It is set at 0.54 according to the measured ratios of $D^+$ and $D^0$, $D^{*+}$ and $D^{*0}$,  $D^+$ and $D^{*+}$, $D^0$ and $D^{*0}$~\cite{Lisovyi:2015uqa}, instead of its default value of 0.75. The details of the simulation can be found in the Supplemental Material. The resulting yields of primary hadrons are found in reasonable agreement with the experimental data as shown in Table~\ref{tab:table5}. We stress that this level of agreement with the data (about 10\%) is enough for our purpose of estimating the production yields of $D_{s0}(2317)$, $D_{s1}(2460)$, and $K_{c\bar{c}}(4180)$ and therefore we do not further fine-tune the PACIAE parameters.

From the produced primary hadrons, we can calculate the production yields of $D_{s0}^*(2317)$ and $D_{s1}(2460)$ using the wave functions given in Eq.~(3) with the Wigner function approach. The predicted yields are given in Table~\ref{tab:table6} in comparison with the BABAR measurements,  where Case 1, 2, and 3 correspond to the results obtained with the three wave functions for different interaction ranges of 1, 2, and 3 fm (see the Supplemental Material for details). The uncertainties in Table~\ref{tab:table6} are obtained by varying the simulation parameter $\rm r_p$ from its reference value of 1.16~\footnote{This value reflects the size of a typical hadron,  such as that of the nucleon, which is about 1 fm, and it was determined by reproducing the yields of light (anti-)nuclei in $pp$ collisions\cite{Yan:PACIAE_light_nuclei}.} by 20\% .  This parameter sets the radius of the sphere centered at the position of a parent particle, where the daughter particles are located, and which affects the dispersion of final states in phase space and controls the hadron rescattering effect considered in the PACIAE model.  As a result, it can affect the production yields of composite particles, such as those studied here. We find again very reasonable agreement between the theoretical yields and the experimental measurements. In particular, the ratio  $R=Y_{D_{s0}^*(2317)}/Y_{D_{s1}(2460)}$  is found to be about 1.5,  also in reasonable agreement with data. The agreements in terms of both the absolute production yields and relative ratio provide a highly nontrivial support for the molecular nature of $D_{s0}^*(2317)$ and $D_{s1}(2460)$. We stress that it is the first time that the experimental measurements have been reproduced. In addition, we note that the production yields of $D_{s0}^*(2317)$/$D_{s1}(2460)$ decrease with the increasing size of the molecules from Case 1 to Case 3, while the ratio stays almost constant, which is a manifestation of the underlying heavy-quark spin symmetry relating the $DK$ and $D^*K$ interactions.  We stress that  the production of hadrons in electron-position collisions is a very involved process, therefore we think that the level of agreement obtained in this work, taking into account  the theoretical and experimental uncertainties, is reasonable. Clearly, more accurate data will undoubtedly further refine our knowledge on these enigmatic mesons.

It is interesting to check whether the conventional $c\bar{s}$ picture for $D_{s0}^*(2317)$ and $D_{s1}(2460)$ can explain the BABAR data. For this,  we turn to the statistical model~\cite{Becattini:1995if}, the details of which can be found in the Supplemental Material. The corresponding results are shown in Table~\ref{tab:dmesons2}. We note that although the absolute production yields are in the ballpark of $10^{-3}$ (consistent with the data), the ratio $R=Y_{D_{s0}^*(2317)}/Y_{D_{s1}(2460)}$ is about $0.8$, much different from the experimental central value but marginally consistent with the lower bound considering the large experimental uncertainty.~\footnote{If we replace the $D_{s0}^*(2317)$ and $D_{s1}(2460)$ masses with those of the GI model~\cite{Godfrey:1985xj}, the ratio will become 0.3, much smaller than the experimental value, which shows again the inadequacy of the GI model for these two states.} We stress that the ratio is a very robust prediction of the statistical model, where the production yield of a vector $D_s$ meson is larger than its pseudoscalar cousin mainly by the spin factor, as is the case for $D^*$ and $D$ mesons  (see the Supplemental Material for details). As a result, the ratio of the production yields of $D_{s0}^*(2317)$ and $D_{s1}(2460)$ provides strong and nontrivial support for their molecular nature as $DK$ and $D^*K$ bound states. For reference, we also calculated the production yield of $D_{s1}(2536)$, which is a typical excited $c\bar{s}$ state~\cite{Yang:2021tvc}, and the result is consistent with the experimental measurement at the level of $20 \sim30\%$ as expected. 

\begin{table*}[htbp] \caption{\label{tab:dmesons2}Charm fragmentation production of $D_{s0}^*(2317)$, $D_{s1}(2460)$, and $D_{s1}(2536)$ in the statistical model. The masses of the $D_{s0}^*(2317)$ and $D_{s1}(2460)$, and $D_{s1}(2536)$ are taken from the review of particle physics~\cite{Workman:2022ynf}, and the results for $D_{s0}^*(2317)$ and $D_{s1}(2460)$ are obtained for the momentum range $p^*>3.2$ GeV/c.}
 \setlength{\tabcolsep}{3.2pt}
 \begin{tabular}{ccccccc}
 \hline
 \hline
 $f(c \to D_s)$& Statistical model &BABAR~\cite{BaBar:2006eep}&ALEPH~\cite{ALEPH:2001fud}&ZEUS~\cite{ZEUS:2008nzg}\\ \hline
$D_{s0}^*(2317)$&$3.6\times 10^{-3}$&$5.37\times 10^{-3}$&$-$&$-$\\ 
$D_{s1}(2460)$&$4.7\times 10^{-3}$&$3.91\times 10^{-3}$&$-$&$-$\\
$D_{s1}(2536)$&$7.5\times 10^{-3}$&$-$&$(9.4\pm 2.2 \pm 0.7)\times10^{-3}$&$(11.1\pm 1.6 ^{+ 0.8}_{-1.0})\times10^{-3}$\\
\hline
 \hline
\end{tabular}
\end{table*}

\begin{figure}[htpb]
\centering
\includegraphics[width=0.45\textwidth]{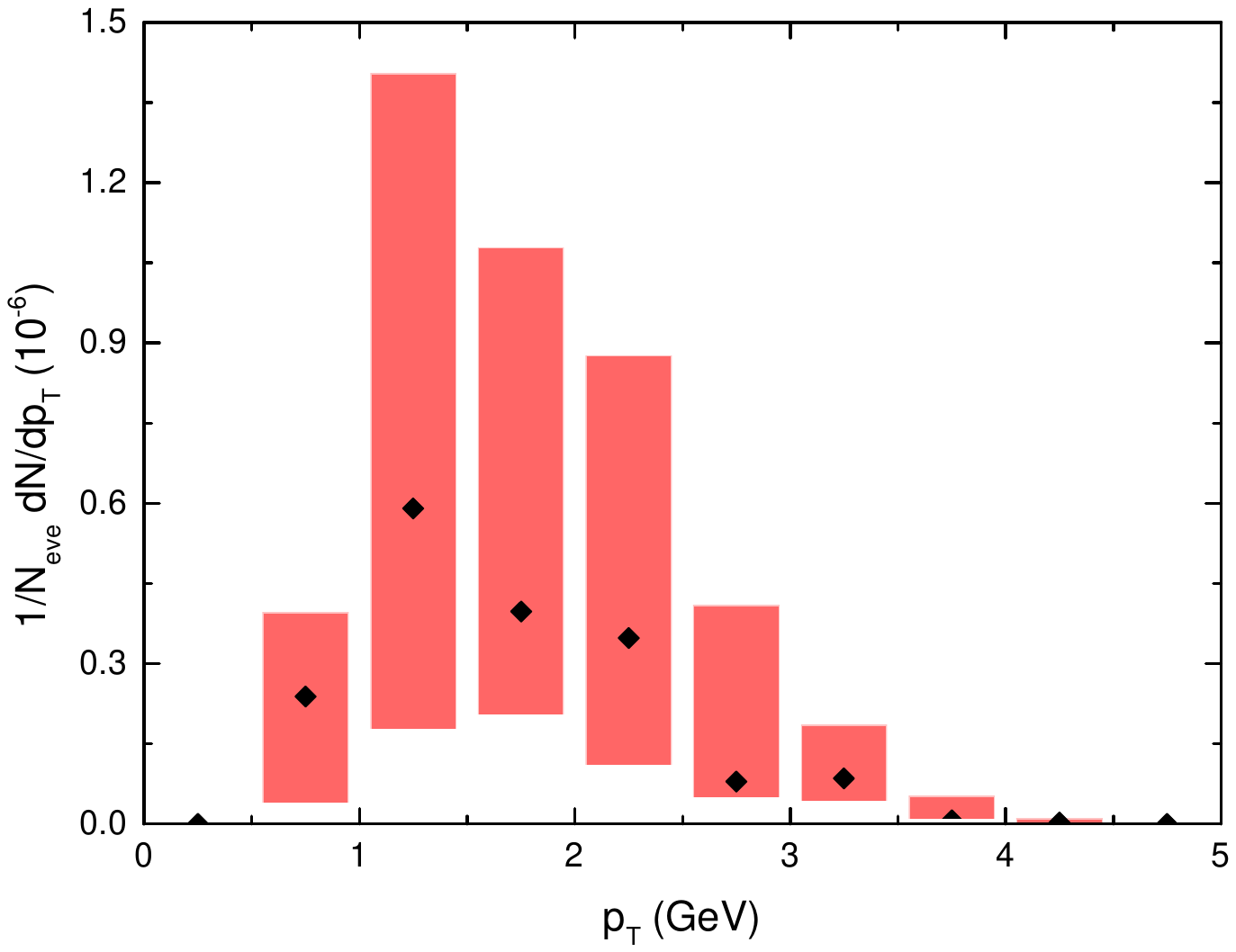}
\caption{ \label{fig:ptdistribution } Transverse momentum distribution of the $D\bar{D}K$ yield in bins of 0.5 GeV, where the uncertainty bands in red are generated by varying $\rm r_p$ from its default value by 20\%.}
\end{figure}

\begin{figure}[htpb]
\centering
\includegraphics[width=0.45\textwidth]{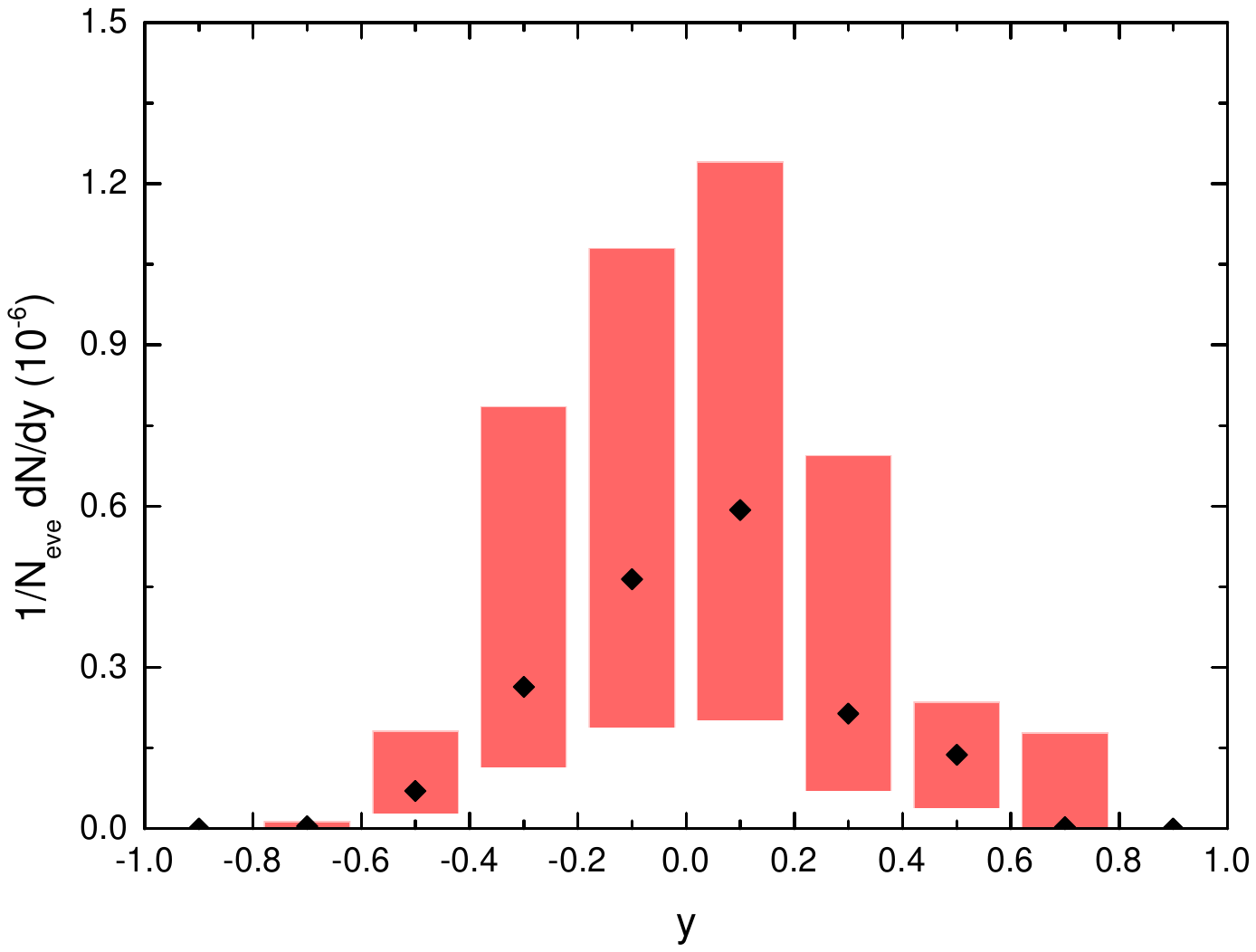}
\caption{ \label{fig:ydistribution } Rapidity distribution of the $D\bar{D}K$ yield in bins of 0.2, where the uncertainty bands in red are generated by varying $r_p$ from its default value by 20\%.}
\end{figure}

Having verified the validity of our transport plus coalescence model, we now study the $DD\bar{K}$ molecule in the same framework. 
Considering all the three Jacobi channels of the $D\bar{D}K$ system~\cite{Wu:2021dwy}, the yield of $D\bar{D}K$ per $e^+e^-\to c\bar{c}$ event (containing charge conjugated states) is found to be $1.75^{+2.66}_{-1.11}\times 10^{-6}$, which is three orders of magnitude lower than the yields of $D_{s0}^*(2317)$ and $D_{s1}(2460)$. Such a reduction in the production yield of a three-body bound state in comparison with that of a two-body bound state is consistent with those observed for deuteron and triton~\cite{ALICE:2017dt3Hebar,ALICE:2021d3Hemultiplicity}. To facilitate experimental searches, we show the transverse momentum and rapidity distribution of $D\bar{D}K$  in Fig.~\ref{fig:ptdistribution }and Fig.~\ref{fig:ydistribution }. We note that the spectra are similar to those of normal hadrons.  

We can now estimate the number of $K_{c\bar{c}}(4180)$ expected in $e^+e^-$ collisions. In Ref.~\cite{BaBar:2006eep}, the number of $D_{s0}^*(2317)$ observed in the $D_s\pi$ mode is $26 290\pm 650$, therefore, the number of $K_{c\bar{c}}(4180)$ produced can be estimated as $N_{D_{s0}^*(2317)}\times Y_{K_{c\bar{c}}(4180)}/Y_{D_{s0}^*(2317)}\approx 10$. In Ref.~\cite{Wu:2020job}, it was shown that $K_{c\bar{c}}(4180)$ decays dominantly to $J/\psi K$. Both final states can be easily measured in $e^+e^-$ collisions~\cite{Belle:2014fgf}. Considering further that the Belle experiment accumulated a data sample of 980 fb$^{-1}$, which is about four times larger than that of BABAR, 232 fb$^{-1}$~\cite{BaBar:2006eep}, we estimate that the Belle data might contain as much as one hundred $K_{c\bar{c}}(4180)$. Belle II will collect 50 ab$^{-1}$ data~\cite{Belle-II:2018jsg} and therefore the number of $K_{c\bar{c}}(4180)$ can reach the order of a few thousands.

{ \it Summary and Outlook:}
In this work, assuming $D_{s0}^*(2317)$ and $D_{s1}(2460)$ as $DK$ and $D^{\ast}K$  molecules, respectively, we investigated  the prompt production yields of $D_{s0}^*(2317)$ and $D_{s1}(2460)$ in inclusive $e^{+}e^{-}\to c\bar{c}$ collisions.  The productions of primary hadrons ($D^{+}$, $D^{+\ast}$, and $K^{-}$) were simulated in the transport model PACIAE, and the formations of $DK$, $D^*K$, and $D\bar{D}K$ bound states were estimated in the Wigner function approach with the wave functions from the accurate Gaussian expansion method.
We find that the prompt production yields of $D_{s0}^*(2317)$ and $D_{s1}(2460)$, and their ratio, which is less sensitive to theoretical uncertainties, are in nice agreement with the available data. These results provide non-trivial evidence  that $D_{s0}^*(2317)$  and $D_{s1}(2460)$ are largely $DK$ and $D^*K$ molecules. We further calculated the production yield of the $D\bar{D}K$ state, $K_{c\bar{c}}(4180)$, and found that it is of the order of $10^{-6}$. This is within the reach of the ongoing Belle II experiment. As a result we encourage dedicated searches for this exotic state in the near future. We stress that our method can be applied to reveal the nature of other enigmatic hadrons, such as the $\Lambda(1405)$~\cite{Lu:2022hwm}, $X(3872)$~\cite{Belle:2003nnu}, $\bar{K}NN$~\cite{Hyodo:2022xhp}, and $D\bar{D}^*K$~\cite{Ren:2018pcd} states.  Works along this line are in progress.

{\it Acknowledgements:}
 We thank Gang Chen, Benhao Sa, Yuliang Yan, Chengping Shen, Hongge Xu, Sen Jia, and Yang Li for valuable discussions. This work is supported in part by the National Natural Science Foundation of China under Grants No.11975041,  No.11735003, and No.11961141004.

\bibliographystyle{elsarticle-num}
\bibliography{BB.bib}

\begin{thebibliography}{10}
\expandafter\ifx\csname url\endcsname\relax
  \def\url#1{\texttt{#1}}\fi
\expandafter\ifx\csname urlprefix\endcsname\relax\def\urlprefix{URL }\fi
\expandafter\ifx\csname href\endcsname\relax
  \def\href#1#2{#2} \def\path#1{#1}\fi

\bibitem{Belle:2003nnu}
S.~K. Choi, et~al., {Observation of a narrow charmonium-like state in exclusive
  $B^\pm \to K^\pm \pi^+ \pi^- J/\psi$ decays}, Phys. Rev. Lett. 91 (2003)
  262001.
\newblock \href {http://arxiv.org/abs/hep-ex/0309032}
  {\path{arXiv:hep-ex/0309032}}, \href
  {https://doi.org/10.1103/PhysRevLett.91.262001}
  {\path{doi:10.1103/PhysRevLett.91.262001}}.

\bibitem{BaBar:2003oey2317}
B.~Aubert, et~al., {Observation of a narrow meson decaying to $D_s^+ \pi^0$ at
  a mass of 2.32-GeV/c$^2$}, Phys. Rev. Lett. 90 (2003) 242001.
\newblock \href {http://arxiv.org/abs/hep-ex/0304021}
  {\path{arXiv:hep-ex/0304021}}, \href
  {https://doi.org/10.1103/PhysRevLett.90.242001}
  {\path{doi:10.1103/PhysRevLett.90.242001}}.

\bibitem{Gell-Mann:1964ewy}
M.~Gell-Mann, {A Schematic Model of Baryons and Mesons}, Phys. Lett. 8 (1964)
  214--215.
\newblock \href {https://doi.org/10.1016/S0031-9163(64)92001-3}
  {\path{doi:10.1016/S0031-9163(64)92001-3}}.

\bibitem{Zweig:1964jf}
G.~Zweig, {An SU(3) model for strong interaction symmetry and its breaking.
  Version 2}, 1964, pp. 22--101.

\bibitem{Guo:2017jvc}
F.-K. Guo, C.~Hanhart, U.-G. Mei\ss{}ner, Q.~Wang, Q.~Zhao, B.-S. Zou,
  {Hadronic molecules}, Rev. Mod. Phys. 90~(1) (2018) 015004, [Erratum:
  Rev.Mod.Phys. 94, 029901 (2022)].
\newblock \href {http://arxiv.org/abs/1705.00141} {\path{arXiv:1705.00141}},
  \href {https://doi.org/10.1103/RevModPhys.90.015004}
  {\path{doi:10.1103/RevModPhys.90.015004}}.

\bibitem{Olsen:2017bmm}
S.~L. Olsen, T.~Skwarnicki, D.~Zieminska, {Nonstandard heavy mesons and
  baryons: Experimental evidence}, Rev. Mod. Phys. 90~(1) (2018) 015003.
\newblock \href {http://arxiv.org/abs/1708.04012} {\path{arXiv:1708.04012}},
  \href {https://doi.org/10.1103/RevModPhys.90.015003}
  {\path{doi:10.1103/RevModPhys.90.015003}}.

\bibitem{Ali:2017jda}
A.~Ali, J.~S. Lange, S.~Stone, {Exotics: Heavy Pentaquarks and Tetraquarks},
  Prog. Part. Nucl. Phys. 97 (2017) 123--198.
\newblock \href {http://arxiv.org/abs/1706.00610} {\path{arXiv:1706.00610}},
  \href {https://doi.org/10.1016/j.ppnp.2017.08.003}
  {\path{doi:10.1016/j.ppnp.2017.08.003}}.

\bibitem{Brambilla:2019esw}
N.~Brambilla, S.~Eidelman, C.~Hanhart, A.~Nefediev, C.-P. Shen, C.~E. Thomas,
  A.~Vairo, C.-Z. Yuan, {The $XYZ$ states: experimental and theoretical status
  and perspectives}, Phys. Rept. 873 (2020) 1--154.
\newblock \href {http://arxiv.org/abs/1907.07583} {\path{arXiv:1907.07583}},
  \href {https://doi.org/10.1016/j.physrep.2020.05.001}
  {\path{doi:10.1016/j.physrep.2020.05.001}}.

\bibitem{Chen:2021ftn}
S.~Chen, Y.~Li, W.~Qian, Z.~Shen, Y.~Xie, Z.~Yang, L.~Zhang, Y.~Zhang, {Heavy
  Flavour Physics and CP Violation at LHCb: a Ten-Year Review} (11 2021).
\newblock \href {http://arxiv.org/abs/2111.14360} {\path{arXiv:2111.14360}}.

\bibitem{Chen:2022asf}
H.-X. Chen, W.~Chen, X.~Liu, Y.-R. Liu, S.-L. Zhu, {An updated review of the
  new hadron states} (4 2022).
\newblock \href {http://arxiv.org/abs/2204.02649} {\path{arXiv:2204.02649}},
  \href {https://doi.org/10.1088/1361-6633/aca3b6}
  {\path{doi:10.1088/1361-6633/aca3b6}}.

\bibitem{CLEO:2003ggt2317}
D.~Besson, et~al., {Observation of a narrow resonance of mass 2.46-GeV/c**2
  decaying to D*+(s) pi0 and confirmation of the D*(sJ)(2317) state}, Phys.
  Rev. D 68 (2003) 032002, [Erratum: Phys.Rev.D 75, 119908 (2007)].
\newblock \href {http://arxiv.org/abs/hep-ex/0305100}
  {\path{arXiv:hep-ex/0305100}}, \href
  {https://doi.org/10.1103/PhysRevD.68.032002}
  {\path{doi:10.1103/PhysRevD.68.032002}}.

\bibitem{Belle:2003guh2317}
P.~Krokovny, et~al., {Observation of the D(sJ)(2317) and D(sJ)(2457) in B
  decays}, Phys. Rev. Lett. 91 (2003) 262002.
\newblock \href {http://arxiv.org/abs/hep-ex/0308019}
  {\path{arXiv:hep-ex/0308019}}, \href
  {https://doi.org/10.1103/PhysRevLett.91.262002}
  {\path{doi:10.1103/PhysRevLett.91.262002}}.

\bibitem{Godfrey:1985xj}
S.~Godfrey, N.~Isgur, {Mesons in a Relativized Quark Model with
  Chromodynamics}, Phys. Rev. D 32 (1985) 189--231.
\newblock \href {https://doi.org/10.1103/PhysRevD.32.189}
  {\path{doi:10.1103/PhysRevD.32.189}}.

\bibitem{Kolomeitsev:2003molecule}
E.~E. Kolomeitsev, M.~F.~M. Lutz, {On Heavy light meson resonances and chiral
  symmetry}, Phys. Lett. B 582 (2004) 39--48.
\newblock \href {http://arxiv.org/abs/hep-ph/0307133}
  {\path{arXiv:hep-ph/0307133}}, \href
  {https://doi.org/10.1016/j.physletb.2003.10.118}
  {\path{doi:10.1016/j.physletb.2003.10.118}}.

\bibitem{vanBeveren:2003molecule}
E.~van Beveren, G.~Rupp, {Observed $D_s(2317)$ and tentative
  $D(2100\text{--}2300)$ as the charmed cousins of the light scalar nonet},
  Phys. Rev. Lett. 91 (2003) 012003.
\newblock \href {http://arxiv.org/abs/hep-ph/0305035}
  {\path{arXiv:hep-ph/0305035}}, \href
  {https://doi.org/10.1103/PhysRevLett.91.012003}
  {\path{doi:10.1103/PhysRevLett.91.012003}}.

\bibitem{Barnes:2003molecule}
T.~Barnes, F.~E. Close, H.~J. Lipkin, {Implications of a DK molecule at
  2.32-GeV}, Phys. Rev. D 68 (2003) 054006.
\newblock \href {http://arxiv.org/abs/hep-ph/0305025}
  {\path{arXiv:hep-ph/0305025}}, \href
  {https://doi.org/10.1103/PhysRevD.68.054006}
  {\path{doi:10.1103/PhysRevD.68.054006}}.

\bibitem{Chen:2004molecule}
Y.-Q. Chen, X.-Q. Li, {A Comprehensive four-quark interpretation of D(s)(2317),
  D(s)(2457) and D(s)(2632)}, Phys. Rev. Lett. 93 (2004) 232001.
\newblock \href {http://arxiv.org/abs/hep-ph/0407062}
  {\path{arXiv:hep-ph/0407062}}, \href
  {https://doi.org/10.1103/PhysRevLett.93.232001}
  {\path{doi:10.1103/PhysRevLett.93.232001}}.

\bibitem{Guo:2006molecule2}
F.-K. Guo, P.-N. Shen, H.-C. Chiang, {Dynamically generated 1+ heavy mesons},
  Phys. Lett. B 647 (2007) 133--139.
\newblock \href {http://arxiv.org/abs/hep-ph/0610008}
  {\path{arXiv:hep-ph/0610008}}, \href
  {https://doi.org/10.1016/j.physletb.2007.01.050}
  {\path{doi:10.1016/j.physletb.2007.01.050}}.

\bibitem{Liu:2012lattice}
L.~Liu, K.~Orginos, F.-K. Guo, C.~Hanhart, U.-G. Meissner, {Interactions of
  charmed mesons with light pseudoscalar mesons from lattice QCD and
  implications on the nature of the $D_{s0}^*(2317)$}, Phys. Rev. D 87~(1)
  (2013) 014508.
\newblock \href {http://arxiv.org/abs/1208.4535} {\path{arXiv:1208.4535}},
  \href {https://doi.org/10.1103/PhysRevD.87.014508}
  {\path{doi:10.1103/PhysRevD.87.014508}}.

\bibitem{Mohler:2013lattice}
D.~Mohler, C.~B. Lang, L.~Leskovec, S.~Prelovsek, R.~M. Woloshyn,
  {$D_{s0}^*(2317)$ Meson and $D$-Meson-Kaon Scattering from Lattice QCD},
  Phys. Rev. Lett. 111~(22) (2013) 222001.
\newblock \href {http://arxiv.org/abs/1308.3175} {\path{arXiv:1308.3175}},
  \href {https://doi.org/10.1103/PhysRevLett.111.222001}
  {\path{doi:10.1103/PhysRevLett.111.222001}}.

\bibitem{Lang:2014lattice}
C.~B. Lang, L.~Leskovec, D.~Mohler, S.~Prelovsek, R.~M. Woloshyn, {Ds mesons
  with DK and D*K scattering near threshold}, Phys. Rev. D 90~(3) (2014)
  034510.
\newblock \href {http://arxiv.org/abs/1403.8103} {\path{arXiv:1403.8103}},
  \href {https://doi.org/10.1103/PhysRevD.90.034510}
  {\path{doi:10.1103/PhysRevD.90.034510}}.

\bibitem{Altenbuchinger:2013vwa}
M.~Altenbuchinger, L.~S. Geng, W.~Weise, {Scattering lengths of Nambu-Goldstone
  bosons off $D$ mesons and dynamically generated heavy-light mesons}, Phys.
  Rev. D 89~(1) (2014) 014026.
\newblock \href {http://arxiv.org/abs/1309.4743} {\path{arXiv:1309.4743}},
  \href {https://doi.org/10.1103/PhysRevD.89.014026}
  {\path{doi:10.1103/PhysRevD.89.014026}}.

\bibitem{MartinezTorres:2014kpc}
A.~Mart\'\i{}nez~Torres, E.~Oset, S.~Prelovsek, A.~Ramos, {Reanalysis of
  lattice QCD spectra leading to the $D_{s0}^*(2317)$ and $D_{s1}^*(2460)$},
  JHEP 05 (2015) 153.
\newblock \href {http://arxiv.org/abs/1412.1706} {\path{arXiv:1412.1706}},
  \href {https://doi.org/10.1007/JHEP05(2015)153}
  {\path{doi:10.1007/JHEP05(2015)153}}.

\bibitem{Yang:2021tvc}
Z.~Yang, G.-J. Wang, J.-J. Wu, M.~Oka, S.-L. Zhu, {Novel Coupled Channel
  Framework Connecting the Quark Model and Lattice QCD for the Near-threshold
  Ds States}, Phys. Rev. Lett. 128~(11) (2022) 112001.
\newblock \href {http://arxiv.org/abs/2107.04860} {\path{arXiv:2107.04860}},
  \href {https://doi.org/10.1103/PhysRevLett.128.112001}
  {\path{doi:10.1103/PhysRevLett.128.112001}}.

\bibitem{Faessler:2007cu}
A.~Faessler, T.~Gutsche, S.~Kovalenko, V.~E. Lyubovitskij, {D*(s0)(2317) and
  D(s1)(2460) mesons in two-body B-meson decays}, Phys. Rev. D 76 (2007)
  014003.
\newblock \href {http://arxiv.org/abs/0705.0892} {\path{arXiv:0705.0892}},
  \href {https://doi.org/10.1103/PhysRevD.76.014003}
  {\path{doi:10.1103/PhysRevD.76.014003}}.

\bibitem{Liu:2022zbd}
M.-Z. Liu, X.-Z. Ling, L.-S. Geng, En-Wang, J.-J. Xie, {Production of
  $D^*_{s0}(2317)$ and $D_{s1}(2460)$ in $B$ decays as $D^{(*)}K$ and
  $D^{(*)}_s\eta$ molecules} (9 2022).
\newblock \href {http://arxiv.org/abs/2209.01103} {\path{arXiv:2209.01103}}.

\bibitem{Dai:2003yg}
Y.-B. Dai, C.-S. Huang, C.~Liu, S.-L. Zhu, {Understanding the D+(sJ)(2317) and
  D+(sJ)(2460) with sum rules in HQET}, Phys. Rev. D 68 (2003) 114011.
\newblock \href {http://arxiv.org/abs/hep-ph/0306274}
  {\path{arXiv:hep-ph/0306274}}, \href
  {https://doi.org/10.1103/PhysRevD.68.114011}
  {\path{doi:10.1103/PhysRevD.68.114011}}.

\bibitem{Lakhina:2006fy}
O.~Lakhina, E.~S. Swanson, {A Canonical Ds(2317)?}, Phys. Lett. B 650 (2007)
  159--165.
\newblock \href {http://arxiv.org/abs/hep-ph/0608011}
  {\path{arXiv:hep-ph/0608011}}, \href
  {https://doi.org/10.1016/j.physletb.2007.01.075}
  {\path{doi:10.1016/j.physletb.2007.01.075}}.

\bibitem{Barnes:2003dj}
T.~Barnes, F.~E. Close, H.~J. Lipkin, {Implications of a DK molecule at
  2.32-GeV}, Phys. Rev. D 68 (2003) 054006.
\newblock \href {http://arxiv.org/abs/hep-ph/0305025}
  {\path{arXiv:hep-ph/0305025}}, \href
  {https://doi.org/10.1103/PhysRevD.68.054006}
  {\path{doi:10.1103/PhysRevD.68.054006}}.

\bibitem{Browder:2003fk}
T.~E. Browder, S.~Pakvasa, A.~A. Petrov, {Comment on the new D(s)(*)+ pi0
  resonances}, Phys. Lett. B 578 (2004) 365--368.
\newblock \href {http://arxiv.org/abs/hep-ph/0307054}
  {\path{arXiv:hep-ph/0307054}}, \href
  {https://doi.org/10.1016/j.physletb.2003.10.067}
  {\path{doi:10.1016/j.physletb.2003.10.067}}.

\bibitem{Wu:2019vsy}
T.-W. Wu, M.-Z. Liu, L.-S. Geng, E.~Hiyama, M.~P. Valderrama, {$DK$, $DDK$, and
  $DDDK$ molecules\textendash{}understanding the nature of the
  $D_{s0}^*(2317)$}, Phys. Rev. D 100~(3) (2019) 034029.
\newblock \href {http://arxiv.org/abs/1906.11995} {\path{arXiv:1906.11995}},
  \href {https://doi.org/10.1103/PhysRevD.100.034029}
  {\path{doi:10.1103/PhysRevD.100.034029}}.

\bibitem{MartinezTorres:2018zbl}
A.~Martinez~Torres, K.~P. Khemchandani, L.-S. Geng, {Bound state formation in
  the $DDK$ system}, Phys. Rev. D 99~(7) (2019) 076017.
\newblock \href {http://arxiv.org/abs/1809.01059} {\path{arXiv:1809.01059}},
  \href {https://doi.org/10.1103/PhysRevD.99.076017}
  {\path{doi:10.1103/PhysRevD.99.076017}}.

\bibitem{Ren:2018pcd}
X.-L. Ren, B.~B. Malabarba, L.-S. Geng, K.~P. Khemchandani,
  A.~Mart\'\i{}nez~Torres, {$K^*$ mesons with hidden charm arising from
  $KX(3872)$ and $KZ_c(3900)$ dynamics}, Phys. Lett. B 785 (2018) 112--117.
\newblock \href {http://arxiv.org/abs/1805.08330} {\path{arXiv:1805.08330}},
  \href {https://doi.org/10.1016/j.physletb.2018.08.034}
  {\path{doi:10.1016/j.physletb.2018.08.034}}.

\bibitem{Wu:2021dwy}
T.-W. Wu, M.-Z. Liu, L.-S. Geng, {One Way to Verify the Molecular Picture of
  Exotic Hadrons: From $\pmb {DK}$ to $\pmb {DDK/D{\bar{D}}^{(*)}K}$}, Few Body
  Syst. 62~(3) (2021) 38.
\newblock \href {http://arxiv.org/abs/2105.09017} {\path{arXiv:2105.09017}},
  \href {https://doi.org/10.1007/s00601-021-01619-y}
  {\path{doi:10.1007/s00601-021-01619-y}}.

\bibitem{MartinezTorres:2020hus}
A.~Martinez~Torres, K.~P. Khemchandani, L.~Roca, E.~Oset, {Few-body systems
  consisting of mesons}, Few Body Syst. 61~(4) (2020) 35.
\newblock \href {http://arxiv.org/abs/2005.14357} {\path{arXiv:2005.14357}},
  \href {https://doi.org/10.1007/s00601-020-01568-y}
  {\path{doi:10.1007/s00601-020-01568-y}}.

\bibitem{Wu:2022ftm}
T.-W. Wu, Y.-W. Pan, M.-Z. Liu, L.-S. Geng, {Multi-hadron molecules: status and
  prospect}, Sci. Bull. 67 (2022) 1735--1738.
\newblock \href {http://arxiv.org/abs/2208.00882} {\path{arXiv:2208.00882}},
  \href {https://doi.org/10.1016/j.scib.2022.08.007}
  {\path{doi:10.1016/j.scib.2022.08.007}}.

\bibitem{Wu:2021gyn}
T.-W. Wu, Y.-W. Pan, M.-Z. Liu, J.-X. Lu, L.-S. Geng, X.-H. Liu, {Hidden charm
  hadronic molecule with strangeness Pcs*(4739) as a
  \ensuremath{\Sigma}cD\textasciimacron{}K\textasciimacron{} bound state},
  Phys. Rev. D 104~(9) (2021) 094032.
\newblock \href {http://arxiv.org/abs/2106.11450} {\path{arXiv:2106.11450}},
  \href {https://doi.org/10.1103/PhysRevD.104.094032}
  {\path{doi:10.1103/PhysRevD.104.094032}}.

\bibitem{Wu:2020job}
T.-W. Wu, M.-Z. Liu, L.-S. Geng, {Excited $K$ meson, $K_c(4180)$ , with hidden
  charm as a $D\bar D K$ bound state}, Phys. Rev. D 103~(3) (2021) L031501.
\newblock \href {http://arxiv.org/abs/2012.01134} {\path{arXiv:2012.01134}},
  \href {https://doi.org/10.1103/PhysRevD.103.L031501}
  {\path{doi:10.1103/PhysRevD.103.L031501}}.

\bibitem{Wu:2020rdg}
T.-W. Wu, M.-Z. Liu, L.-S. Geng, E.~Hiyama, M.~P. Valderrama, W.-L. Wang,
  {Quadruply charmed dibaryons as heavy quark symmetry partners of the $DDK$
  bound state}, Eur. Phys. J. C 80~(9) (2020) 901.
\newblock \href {http://arxiv.org/abs/2004.09779} {\path{arXiv:2004.09779}},
  \href {https://doi.org/10.1140/epjc/s10052-020-08483-w}
  {\path{doi:10.1140/epjc/s10052-020-08483-w}}.

\bibitem{Wu:2021kbu}
T.-W. Wu, Y.-W. Pan, M.-Z. Liu, S.-Q. Luo, L.-S. Geng, X.~Liu, {Discovery of
  the doubly charmed Tcc+ state implies a triply charmed Hccc hexaquark state},
  Phys. Rev. D 105~(3) (2022) L031505.
\newblock \href {http://arxiv.org/abs/2108.00923} {\path{arXiv:2108.00923}},
  \href {https://doi.org/10.1103/PhysRevD.105.L031505}
  {\path{doi:10.1103/PhysRevD.105.L031505}}.

\bibitem{Luo:2021ggs}
S.-Q. Luo, T.-W. Wu, M.-Z. Liu, L.-S. Geng, X.~Liu, {Triple-charm molecular
  states composed of D*D*D and D*D*D*}, Phys. Rev. D 105~(7) (2022) 074033.
\newblock \href {http://arxiv.org/abs/2111.15079} {\path{arXiv:2111.15079}},
  \href {https://doi.org/10.1103/PhysRevD.105.074033}
  {\path{doi:10.1103/PhysRevD.105.074033}}.

\bibitem{Pan:2022xxz}
Y.-W. Pan, T.-W. Wu, M.-Z. Liu, L.-S. Geng, {Three-body molecules
  D\textasciimacron{}D\textasciimacron{}*\ensuremath{\Sigma}c: Understanding
  the nature of Tcc, Pc(4312), Pc(4440), and Pc(4457)}, Phys. Rev. D 105~(11)
  (2022) 114048.
\newblock \href {http://arxiv.org/abs/2204.02295} {\path{arXiv:2204.02295}},
  \href {https://doi.org/10.1103/PhysRevD.105.114048}
  {\path{doi:10.1103/PhysRevD.105.114048}}.

\bibitem{Belle:2020xca}
Y.~Li, et~al., {Search for a doubly-charged $DDK$ bound state in
  $\Upsilon(1S,2S)$ inclusive decays and via direct production in $e^+e^-$
  collisions at $\sqrt{s}$ = 10.520, 10.580, and 10.867 GeV}, Phys. Rev. D
  102~(11) (2020) 112001.
\newblock \href {http://arxiv.org/abs/2008.13341} {\path{arXiv:2008.13341}},
  \href {https://doi.org/10.1103/PhysRevD.102.112001}
  {\path{doi:10.1103/PhysRevD.102.112001}}.

\bibitem{Mattiello:1996wignertheory3}
R.~Mattiello, H.~Sorge, H.~Stoecker, W.~Greiner, {Nuclear clusters as a probe
  for expansion flow in heavy ion reactions at 10-A/GeV - 15-A/GeV}, Phys. Rev.
  C 55 (1997) 1443--1454.
\newblock \href {http://arxiv.org/abs/nucl-th/9607003}
  {\path{arXiv:nucl-th/9607003}}, \href
  {https://doi.org/10.1103/PhysRevC.55.1443}
  {\path{doi:10.1103/PhysRevC.55.1443}}.

\bibitem{Nagle:1996cutoff4Wigner}
J.~L. Nagle, B.~S. Kumar, D.~Kusnezov, H.~Sorge, R.~Mattiello, {Coalescence of
  deuterons in relativistic heavy ion collisions}, Phys. Rev. C 53 (1996)
  367--376.
\newblock \href {https://doi.org/10.1103/PhysRevC.53.367}
  {\path{doi:10.1103/PhysRevC.53.367}}.

\bibitem{STAR:2011eej}
H.~Agakishiev, et~al., {Observation of the antimatter helium-4 nucleus}, Nature
  473 (2011) 353, [Erratum: Nature 475, 412 (2011)].
\newblock \href {http://arxiv.org/abs/1103.3312} {\path{arXiv:1103.3312}},
  \href {https://doi.org/10.1038/nature10079} {\path{doi:10.1038/nature10079}}.

\bibitem{ALICE:2022jmr}
{Enhanced deuteron coalescence probability in jets} (11 2022).
\newblock \href {http://arxiv.org/abs/2211.15204} {\path{arXiv:2211.15204}}.

\bibitem{STAR:2010gyg}
B.~I. Abelev, et~al., {Observation of an Antimatter Hypernucleus}, Science 328
  (2010) 58--62.
\newblock \href {http://arxiv.org/abs/1003.2030} {\path{arXiv:1003.2030}},
  \href {https://doi.org/10.1126/science.1183980}
  {\path{doi:10.1126/science.1183980}}.

\bibitem{Gev:2022ksw}
Gev, {First Observation of Directed Flow of Hypernuclei $^3_{\Lambda}$H and
  $^4_{\Lambda}$H in $\sqrt{s_{\rm NN}}$ = 3 GeV Au+Au Collisions at RHIC} (11
  2022).
\newblock \href {http://arxiv.org/abs/2211.16981} {\path{arXiv:2211.16981}}.

\bibitem{Exotic:2010relativistic5coalescence}
S.~Cho, et~al., {Multi-quark hadrons from Heavy Ion Collisions}, Phys. Rev.
  Lett. 106 (2011) 212001.
\newblock \href {http://arxiv.org/abs/1011.0852} {\path{arXiv:1011.0852}},
  \href {https://doi.org/10.1103/PhysRevLett.106.212001}
  {\path{doi:10.1103/PhysRevLett.106.212001}}.

\bibitem{Zhang:2020dwn}
H.~Zhang, J.~Liao, E.~Wang, Q.~Wang, H.~Xing, {Deciphering the Nature of
  X(3872) in Heavy Ion Collisions}, Phys. Rev. Lett. 126~(1) (2021) 012301.
\newblock \href {http://arxiv.org/abs/2004.00024} {\path{arXiv:2004.00024}},
  \href {https://doi.org/10.1103/PhysRevLett.126.012301}
  {\path{doi:10.1103/PhysRevLett.126.012301}}.

\bibitem{Chen:PACIAE_3872_2022}
C.-h. Chen, Y.-L. Xie, H.-g. Xu, Z.~Zhang, D.-M. Zhou, Z.-L. She, G.~Chen,
  {Exotic states Pc(4312), Pc(4440), and Pc(4457) in pp collisions at s=7,
  13~TeV}, Phys. Rev. D 105~(5) (2022) 054013.
\newblock \href {https://doi.org/10.1103/PhysRevD.105.054013}
  {\path{doi:10.1103/PhysRevD.105.054013}}.

\bibitem{Chen:2021akx}
B.~Chen, L.~Jiang, X.-H. Liu, Y.~Liu, J.~Zhao, {X(3872) production in
  relativistic heavy-ion collisions}, Phys. Rev. C 105~(5) (2022) 054901.
\newblock \href {http://arxiv.org/abs/2107.00969} {\path{arXiv:2107.00969}},
  \href {https://doi.org/10.1103/PhysRevC.105.054901}
  {\path{doi:10.1103/PhysRevC.105.054901}}.

\bibitem{Hu:2021gdg}
Y.~Hu, J.~Liao, E.~Wang, Q.~Wang, H.~Xing, H.~Zhang, {Production of doubly
  charmed exotic hadrons in heavy ion collisions}, Phys. Rev. D 104~(11) (2021)
  L111502.
\newblock \href {http://arxiv.org/abs/2109.07733} {\path{arXiv:2109.07733}},
  \href {https://doi.org/10.1103/PhysRevD.104.L111502}
  {\path{doi:10.1103/PhysRevD.104.L111502}}.

\bibitem{Wu:2020zbx}
B.~Wu, X.~Du, M.~Sibila, R.~Rapp, {$X(3872)$transport in heavy-ion collisions},
  Eur. Phys. J. A 57~(4) (2021) 122, [Erratum: Eur.Phys.J.A 57, 314 (2021)].
\newblock \href {http://arxiv.org/abs/2006.09945} {\path{arXiv:2006.09945}},
  \href {https://doi.org/10.1140/epja/s10050-021-00623-4}
  {\path{doi:10.1140/epja/s10050-021-00623-4}}.

\bibitem{Abreu:2022lfy}
L.~M. Abreu, H.~P.~L. Vieira, F.~S. Navarra, {Multiplicity of the doubly
  charmed state Tcc+ in heavy-ion collisions}, Phys. Rev. D 105~(11) (2022)
  116029.
\newblock \href {http://arxiv.org/abs/2202.10882} {\path{arXiv:2202.10882}},
  \href {https://doi.org/10.1103/PhysRevD.105.116029}
  {\path{doi:10.1103/PhysRevD.105.116029}}.

\bibitem{Yoon:2022voo}
H.-O. Yoon, D.~Park, S.~Noh, A.~Park, W.~Park, S.~Cho, J.~Hong, Y.~Kim, S.~Lim,
  S.~H. Lee, {$X(3872)$ and $T_{cc}$: structures and productions in heavy ion
  collisions} (8 2022).
\newblock \href {http://arxiv.org/abs/2208.06960} {\path{arXiv:2208.06960}}.

\bibitem{BaBar:2006eep}
B.~Aubert, et~al., {A Study of the D*(sJ)(2317) and D(sJ)(2460) Mesons in
  Inclusive c anti-c Production Near (s)**(1/2) = 10.6-GeV}, Phys. Rev. D 74
  (2006) 032007.
\newblock \href {http://arxiv.org/abs/hep-ex/0604030}
  {\path{arXiv:hep-ex/0604030}}, \href
  {https://doi.org/10.1103/PhysRevD.74.032007}
  {\path{doi:10.1103/PhysRevD.74.032007}}.

\bibitem{Sato:1981ez}
H.~Sato, K.~Yazaki, {On the coalescence model for high-energy nuclear
  reactions}, Phys. Lett. B 98 (1981) 153--157.
\newblock \href {https://doi.org/10.1016/0370-2693(81)90976-X}
  {\path{doi:10.1016/0370-2693(81)90976-X}}.

\bibitem{Sa:2011PACIAE20}
B.-H. Sa, D.-M. Zhou, Y.-L. Yan, X.-M. Li, S.-Q. Feng, B.-G. Dong, X.~Cai,
  {PACIAE 2.0: An Updated parton and hadron cascade model (program) for the
  relativistic nuclear collisions}, Comput. Phys. Commun. 183 (2012) 333--346.
\newblock \href {http://arxiv.org/abs/1104.1238} {\path{arXiv:1104.1238}},
  \href {https://doi.org/10.1016/j.cpc.2011.08.021}
  {\path{doi:10.1016/j.cpc.2011.08.021}}.

\bibitem{Yan:PACIAE_light_nuclei}
Y.-L. Yan, G.~Chen, X.-M. Li, D.-M. Zhou, M.-J. Wang, S.-Y. Hu, L.~Ye, B.-H.
  Sa, {Predictions for the production of light nuclei in $pp$ collisions at
  $\sqrt{s}=7$ and 14 TeV}, Phys. Rev. C 85 (2012) 024907.
\newblock \href {http://arxiv.org/abs/1107.3207} {\path{arXiv:1107.3207}},
  \href {https://doi.org/10.1103/PhysRevC.85.024907}
  {\path{doi:10.1103/PhysRevC.85.024907}}.

\bibitem{Sjostrand:2006PYTHIA}
T.~Sjostrand, S.~Mrenna, P.~Z. Skands, {PYTHIA 6.4 Physics and Manual}, JHEP 05
  (2006) 026.
\newblock \href {http://arxiv.org/abs/hep-ph/0603175}
  {\path{arXiv:hep-ph/0603175}}, \href
  {https://doi.org/10.1088/1126-6708/2006/05/026}
  {\path{doi:10.1088/1126-6708/2006/05/026}}.

\bibitem{Chen:2003deuteronwigner}
L.-W. Chen, C.~M. Ko, B.-A. Li, {Light cluster production in
  intermediate-energy heavy ion collisions induced by neutron rich nuclei},
  Nucl. Phys. A 729 (2003) 809--834.
\newblock \href {http://arxiv.org/abs/nucl-th/0306032}
  {\path{arXiv:nucl-th/0306032}}, \href
  {https://doi.org/10.1016/j.nuclphysa.2003.09.010}
  {\path{doi:10.1016/j.nuclphysa.2003.09.010}}.

\bibitem{Sombun:2018yqh}
S.~Sombun, K.~Tomuang, A.~Limphirat, P.~Hillmann, C.~Herold, J.~Steinheimer,
  Y.~Yan, M.~Bleicher, {Deuteron production from phase-space coalescence in the
  UrQMD approach}, Phys. Rev. C 99~(1) (2019) 014901.
\newblock \href {http://arxiv.org/abs/1805.11509} {\path{arXiv:1805.11509}},
  \href {https://doi.org/10.1103/PhysRevC.99.014901}
  {\path{doi:10.1103/PhysRevC.99.014901}}.

\bibitem{Deng:2020zxo}
X.~G. Deng, Y.~G. Ma, {Light nuclei production in Au + Au collisions at
  $\sqrt{s_{NN}}$ = 7.7-80 GeV from UrQMD model}, Phys. Lett. B 808 (2020)
  135668.
\newblock \href {http://arxiv.org/abs/2006.12337} {\path{arXiv:2006.12337}},
  \href {https://doi.org/10.1016/j.physletb.2020.135668}
  {\path{doi:10.1016/j.physletb.2020.135668}}.

\bibitem{Zhang:2020diabaryon}
S.~Zhang, Y.-G. Ma, {$\Omega$-dibaryon production with hadron interaction
  potential from the lattice QCD in relativistic heavy-ion collisions}, Phys.
  Lett. B 811 (2020) 135867.
\newblock \href {http://arxiv.org/abs/2007.11170} {\path{arXiv:2007.11170}},
  \href {https://doi.org/10.1016/j.physletb.2020.135867}
  {\path{doi:10.1016/j.physletb.2020.135867}}.

\bibitem{Gyulassy:1982wignertheory1}
M.~Gyulassy, K.~Frankel, E.~a. Remler, {DEUTERON FORMATION IN NUCLEAR
  COLLISIONS}, Nucl. Phys. A 402 (1983) 596--611.
\newblock \href {https://doi.org/10.1016/0375-9474(83)90222-1}
  {\path{doi:10.1016/0375-9474(83)90222-1}}.

\bibitem{Kolomeitsev:2003ac}
E.~E. Kolomeitsev, M.~F.~M. Lutz, {On Heavy light meson resonances and chiral
  symmetry}, Phys. Lett. B 582 (2004) 39--48.
\newblock \href {http://arxiv.org/abs/hep-ph/0307133}
  {\path{arXiv:hep-ph/0307133}}, \href
  {https://doi.org/10.1016/j.physletb.2003.10.118}
  {\path{doi:10.1016/j.physletb.2003.10.118}}.

\bibitem{Gamermann:2006nm}
D.~Gamermann, E.~Oset, D.~Strottman, M.~J. Vicente~Vacas, {Dynamically
  generated open and hidden charm meson systems}, Phys. Rev. D 76 (2007)
  074016.
\newblock \href {http://arxiv.org/abs/hep-ph/0612179}
  {\path{arXiv:hep-ph/0612179}}, \href
  {https://doi.org/10.1103/PhysRevD.76.074016}
  {\path{doi:10.1103/PhysRevD.76.074016}}.

\bibitem{Guo:2006fu}
F.-K. Guo, P.-N. Shen, H.-C. Chiang, R.-G. Ping, B.-S. Zou, {Dynamically
  generated 0+ heavy mesons in a heavy chiral unitary approach}, Phys. Lett. B
  641 (2006) 278--285.
\newblock \href {http://arxiv.org/abs/hep-ph/0603072}
  {\path{arXiv:hep-ph/0603072}}, \href
  {https://doi.org/10.1016/j.physletb.2006.08.064}
  {\path{doi:10.1016/j.physletb.2006.08.064}}.

\bibitem{Liu:2012zya}
L.~Liu, K.~Orginos, F.-K. Guo, C.~Hanhart, U.-G. Meissner, {Interactions of
  charmed mesons with light pseudoscalar mesons from lattice QCD and
  implications on the nature of the $D_{s0}^*(2317)$}, Phys. Rev. D 87~(1)
  (2013) 014508.
\newblock \href {http://arxiv.org/abs/1208.4535} {\path{arXiv:1208.4535}},
  \href {https://doi.org/10.1103/PhysRevD.87.014508}
  {\path{doi:10.1103/PhysRevD.87.014508}}.

\bibitem{Guo:2015dha}
Z.-H. Guo, U.-G. Mei\ss{}ner, D.-L. Yao, {New insights into the
  $D^{*}_{s0}(2317)$ and other charm scalar mesons}, Phys. Rev. D 92~(9) (2015)
  094008.
\newblock \href {http://arxiv.org/abs/1507.03123} {\path{arXiv:1507.03123}},
  \href {https://doi.org/10.1103/PhysRevD.92.094008}
  {\path{doi:10.1103/PhysRevD.92.094008}}.

\bibitem{Yao:2015qia}
D.-L. Yao, M.-L. Du, F.-K. Guo, U.-G. Mei\ss{}ner, {One-loop analysis of the
  interactions between charmed mesons and Goldstone bosons}, JHEP 11 (2015)
  058.
\newblock \href {http://arxiv.org/abs/1502.05981} {\path{arXiv:1502.05981}},
  \href {https://doi.org/10.1007/JHEP11(2015)058}
  {\path{doi:10.1007/JHEP11(2015)058}}.

\bibitem{Du:2017ttu}
M.-L. Du, F.-K. Guo, U.-G. Mei\ss{}ner, D.-L. Yao, {Study of open-charm $0^+$
  states in unitarized chiral effective theory with one-loop potentials}, Eur.
  Phys. J. C 77~(11) (2017) 728.
\newblock \href {http://arxiv.org/abs/1703.10836} {\path{arXiv:1703.10836}},
  \href {https://doi.org/10.1140/epjc/s10052-017-5287-6}
  {\path{doi:10.1140/epjc/s10052-017-5287-6}}.

\bibitem{Huang:2022cag}
B.-L. Huang, Z.-Y. Lin, K.~Chen, S.-L. Zhu, {Phase shifts of the light
  pseudoscalar meson and heavy meson scattering in heavy meson chiral
  perturbation theory}, Eur. Phys. J. C 83~(1) (2023) 76.
\newblock \href {http://arxiv.org/abs/2205.02619} {\path{arXiv:2205.02619}},
  \href {https://doi.org/10.1140/epjc/s10052-023-11235-1}
  {\path{doi:10.1140/epjc/s10052-023-11235-1}}.

\bibitem{Lisovyi:2015uqa}
M.~Lisovyi, A.~Verbytskyi, O.~Zenaiev, {Combined analysis of charm-quark
  fragmentation-fraction measurements}, Eur. Phys. J. C 76~(7) (2016) 397.
\newblock \href {http://arxiv.org/abs/1509.01061} {\path{arXiv:1509.01061}},
  \href {https://doi.org/10.1140/epjc/s10052-016-4246-y}
  {\path{doi:10.1140/epjc/s10052-016-4246-y}}.

\bibitem{Wei:2022jgc}
X.~Wei, Q.-H. Shen, J.-J. Xie, {Faddeev fixed-center approximation to the
  $D\bar{D}K$ system and the hidden charm $K_{c\bar{c}}(4180)$ state} (5 2022).
\newblock \href {http://arxiv.org/abs/2205.12526} {\path{arXiv:2205.12526}}.

\bibitem{Becattini:1995if}
F.~Becattini, {A Thermodynamical approach to hadron production in e+ e-
  collisions}, Z. Phys. C 69~(3) (1996) 485--492.
\newblock \href {https://doi.org/10.1007/BF02907431}
  {\path{doi:10.1007/BF02907431}}.

\bibitem{Workman:2022ynf}
R.~L. Workman, Others, {Review of Particle Physics}, PTEP 2022 (2022) 083C01.
\newblock \href {https://doi.org/10.1093/ptep/ptac097}
  {\path{doi:10.1093/ptep/ptac097}}.

\bibitem{ALEPH:2001fud}
A.~Heister, et~al., {Production of D**(s) mesons in hadronic Z decays}, Phys.
  Lett. B 526 (2002) 34--49.
\newblock \href {http://arxiv.org/abs/hep-ex/0112010}
  {\path{arXiv:hep-ex/0112010}}, \href
  {https://doi.org/10.1016/S0370-2693(01)01465-4}
  {\path{doi:10.1016/S0370-2693(01)01465-4}}.

\bibitem{ZEUS:2008nzg}
S.~Chekanov, et~al., {Production of excited charm and charm-strange mesons at
  HERA}, Eur. Phys. J. C 60 (2009) 25--45.
\newblock \href {http://arxiv.org/abs/0807.1290} {\path{arXiv:0807.1290}},
  \href {https://doi.org/10.1140/epjc/s10052-009-0881-x}
  {\path{doi:10.1140/epjc/s10052-009-0881-x}}.

\bibitem{ALICE:2017dt3Hebar}
S.~Acharya, et~al., {Production of deuterons, tritons, $^{3}$He nuclei and
  their antinuclei in pp collisions at $\mathbf{\sqrt{{\textit s}}}$ = 0.9,
  2.76 and 7 TeV}, Phys. Rev. C 97~(2) (2018) 024615.
\newblock \href {http://arxiv.org/abs/1709.08522} {\path{arXiv:1709.08522}},
  \href {https://doi.org/10.1103/PhysRevC.97.024615}
  {\path{doi:10.1103/PhysRevC.97.024615}}.

\bibitem{ALICE:2021d3Hemultiplicity}
S.~Acharya, et~al., {Production of light (anti)nuclei in pp collisions at
  $\sqrt{s} = 13$TeV} (9 2021).
\newblock \href {http://arxiv.org/abs/2109.13026} {\path{arXiv:2109.13026}}.

\bibitem{Belle:2014fgf}
C.~P. Shen, et~al., {Updated cross section measurement of $e^+ e^- \to K^+ K^-
  J/\psi$ and $K_S^0K_S^0J/\psi$ via initial state radiation at Belle}, Phys.
  Rev. D 89~(7) (2014) 072015.
\newblock \href {http://arxiv.org/abs/1402.6578} {\path{arXiv:1402.6578}},
  \href {https://doi.org/10.1103/PhysRevD.89.072015}
  {\path{doi:10.1103/PhysRevD.89.072015}}.

\bibitem{Belle-II:2018jsg}
W.~Altmannshofer, et~al., {The Belle II Physics Book}, PTEP 2019~(12) (2019)
  123C01, [Erratum: PTEP 2020, 029201 (2020)].
\newblock \href {http://arxiv.org/abs/1808.10567} {\path{arXiv:1808.10567}},
  \href {https://doi.org/10.1093/ptep/ptz106} {\path{doi:10.1093/ptep/ptz106}}.

\bibitem{Lu:2022hwm}
J.-X. Lu, L.-S. Geng, M.~Doering, M.~Mai, {Cross-channel constraints on
  resonant antikaon-nucleon scattering} (9 2022).
\newblock \href {http://arxiv.org/abs/2209.02471} {\path{arXiv:2209.02471}}.

\bibitem{Hyodo:2022xhp}
T.~Hyodo, W.~Weise, {Theory of kaon-nuclear systems}, 2022.
\newblock \href {http://arxiv.org/abs/2202.06181} {\path{arXiv:2202.06181}}.

\bibitem{WuTianwei:2019vsyDKDDK}
T.-W. Wu, M.-Z. Liu, L.-S. Geng, E.~Hiyama, M.~P. Valderrama, {$DK$, $DDK$, and
  $DDDK$ molecules\textendash{}understanding the nature of the
  $D_{s0}^*(2317)$}, Phys. Rev. D 100~(3) (2019) 034029.
\newblock \href {http://arxiv.org/abs/1906.11995} {\path{arXiv:1906.11995}},
  \href {https://doi.org/10.1103/PhysRevD.100.034029}
  {\path{doi:10.1103/PhysRevD.100.034029}}.

\bibitem{WuTianwei:2021dwyDDbarK}
T.-W. Wu, M.-Z. Liu, L.-S. Geng, {One Way to Verify the Molecular Picture of
  Exotic Hadrons: From from $DK$ to $DDK/D\bar{D}^{(*)}K$}, Few Body Syst.
  62~(3) (2021) 38.
\newblock \href {http://arxiv.org/abs/2105.09017} {\path{arXiv:2105.09017}},
  \href {https://doi.org/10.1007/s00601-021-01619-y}
  {\path{doi:10.1007/s00601-021-01619-y}}.

\bibitem{Liu:2023uly}
Z.-W. Liu, J.-X. Lu, L.-S. Geng, {Study of the $DK$ interaction with
  femtoscopic correlation functions} (2 2023).
\newblock \href {http://arxiv.org/abs/2302.01046} {\path{arXiv:2302.01046}}.

\bibitem{Becattini:2008tx}
F.~Becattini, P.~Castorina, J.~Manninen, H.~Satz, {The Thermal Production of
  Strange and Non-Strange Hadrons in e+ e- Collisions}, Eur. Phys. J. C 56
  (2008) 493--510.
\newblock \href {http://arxiv.org/abs/0805.0964} {\path{arXiv:0805.0964}},
  \href {https://doi.org/10.1140/epjc/s10052-008-0671-x}
  {\path{doi:10.1140/epjc/s10052-008-0671-x}}.

\end{thebibliography}

\clearpage
\begin{widetext}
\setcounter{page}{1}
\section{Supplemental material}
In this Supplemental Material, we provide further details about the wave functions of $D_{s0}^*(2317)$ and $D_{s1}(2460)$, the simulation process, the coalescence model, and the statistical model, which are relevant to understand the results presented in the main text. 
\subsection{Wave functions of $D_{s0}^*(2317)$ and $D_{s1}(2460)$}
The $D_{s0}^*(2317)$ and $D_{s1}(2460)$ can be understood as $DK$ and $D^*K$ molecules bound by the residual strong interaction in the unitary chiral approaches with either leading order~\cite{Guo:2006fu}, next-to-leading order~\cite{Altenbuchinger:2013vwa} or next-to-next-to-leading order~\cite{Du:2017ttu} chiral potentials.  At leading order~\cite{Guo:2006fu}, the Weinberg-Tomozawa term is responsible for the attraction between the $D(D^*)$ and $K$ mesons (and their coupled channels). For most physics related to the $D_{s0}^*(2317)$ and $D_{s1}(2460)$, the leading order chiral potential is enough. The Fourier transform of the Weinberg-Tomozawa potential is a delta function in coordinate space. To take into account the finite sizes of the $D_{s0}^*(2317)/D_{s1}(2460)$ states, the delta function can be approximated with a Gaussian function:
\begin{equation}
V_{D K}\left(r ; R_c\right)=C\left(R_C\right) \frac{e^{-\left(r / R_c\right)^2}}{\pi^{3 / 2} R_c^3},
\end{equation}
 where $R_c$ is a cutoff parameter characterizing the range of the potential and $C(R_c)$ is a running coupling constant determined by fitting to the binding energy of 45 MeV for the $D_{s0}^*(2317)$ as a $DK$ bound state.
 This is what was done in the studies of the $DDK$, $D\bar{D}K$ and $DDDK$ molecules~\cite{WuTianwei:2019vsyDKDDK,WuTianwei:2021dwyDDbarK}, where the wave functions of $D_{s0}^*(2317)$ and $D_{s1}(2460)$ are obtained by solving the Schroedinger equation using the Gaussian Expansion Method (GEM). In the GEM, the wave function is expanded by $N$ Gaussian bases with different widths parameterized by $\omega_i$. We show in Fig.~\ref{fig:dk3wavefunction} the wave function of the $D_{s0}^*(2317)$ as a function of the radial distance $r$ between $D$ and $K$ obtained with a cutoff of $R_c=1,2,3$ fm. These wave functions yield a root-mean-square radius of 1.28, 1.74, and 2.13 fm for the $D_{s0}^*(2317)$\cite{WuTianwei:2019vsyDKDDK}. Future experimental measurements of such a quantity will help fix the cutoff $R_c$. In Ref.~\cite{Liu:2023uly}, the $DK$ femtoscopic correlation function was computed, which if measured by future experiments, will also help better determine the $DK$ interaction. 
\begin{figure}[htpb]
\centering
\includegraphics[width=0.48\textwidth]{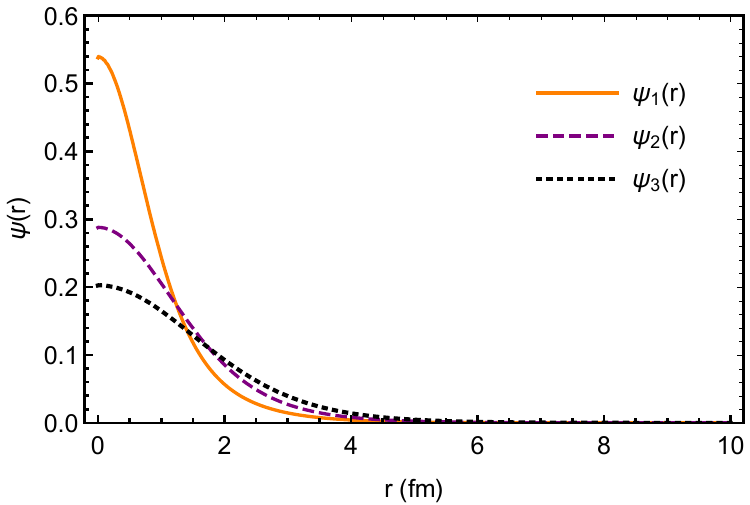}
\caption{ \label{fig:dk3wavefunction}  Wave function of $D_{s0}^*(2317)$ obtained with three different cutoffs of $R_c=1,2,3$ fm.}
\end{figure}
\subsection{Simulation details}
The PACIAE model is a transport model based on the event generator PYTHIA~\cite{Sjostrand:2006PYTHIA}. The model can simulate several types of high-energy collisions between leptons, protons, and nuclei. It simulates these collisions in four stages~\cite{Sa:2011PACIAE20}: Parton initiation, parton rescattering, hadronization, and hadron rescattering. Parton initiation and hadronization are the same as those in  PYTHIA. In parton initiation, hadron-hadron collisions are decomposed further into parton-parton interactions, in which the hard process is calculated by the leading order perturbative QCD parton-parton interactions, while the non-perturbative part is described empirically. In PYTHIA, this process is followed by hadronization and hadron decay directly. Hadronization is described by the LUND string model, where the color field between a quark pair is described by a one-dimensional string. The excited strings are always fragmented into lower-energy strings until they are stable, which corresponds to the formation of hadrons. Via tunneling, the probability of creating quark pairs of different flavors and momentum from the vacuum at the fragmentation point is related to the masses of different quarks and fragmentation functions. The
PACIAE model, however, introduces additional transport processes before and after hadronization, i.e. parton rescattering and hadron rescattering. The former describes the dynamic properties of QGP (if it is produced), while the latter indicates that hadrons interact with each other until they reach the kinetic equilibrium after hadronization. The transport of partons and hadrons affects their final phase space distribution. 

In this work, all the parameters of the PACIAE model are kept at their default values, which are determined by reproducing the LEP $e^+e^- \to Z^0$ data at around 91 GeV~\cite{Sjostrand:2006PYTHIA}, except for the parameter PARJ(13), the probability that a charm or heavier meson has spin 1~\cite{Sjostrand:2006PYTHIA}. In the rescattering stages, since there are almost no parton interactions in the $e^+e^- \to c\bar{c}$ process, we only take into account the effect of hadron rescattering after hadronization in final phase space distributions. Thus we also study the impact of different $r_p$ values on the final simulation results, which affect the relative distance of parimary particles and therefore final-state rescattering. In our simulation, we used $10^6$ events to study the two-body $DK$ and $D^*K$ systems, and $10^8$ events to study the three-body $D\bar{D}K$ system to obtain stable simulation results.

\subsection{Coalescence model}
The coalescence model embodies the dynamic cascade process, because only when the interactions between hadrons almost cease and the temperature of the system is lower to reach kinetic equilibrium, can hadrons combine into relatively stable composite particles. This method was first used to study the production yields of nuclear clusters in high-energy nucleus-nucleus collisions\cite{Sato:1981ez,Gyulassy:1982wignertheory1}. Particles in the neighborhood of the bound pairs of interest are assumed to always absorb their binding energy by interactions to satisfy the energy conservation of the whole system. The yield described by the overlap of the density of the composite particle and the density of the particle source is always larger than the actual yield because it contains the contribution of other many-body subsystems formed from the particle source that have the component of this composite particle. In the $e^+e^- \to c\bar{c} \to$ hadrons process, the probability of a many-body subsystem containing $DK$  is very small, so it is reasonable to neglect the overestimated part, and it is the same for the three-body system $D\bar{D}K$. The interactions, described by instant collisions between which particles travel in free straight trajectories, can be divided into two parts in chronological order, i.e., intermediate interactions, and final interaction. It is reasonable to assume that the intermediate interactions are largely canceled out, and the production rate can be approximated by the final state contribution. Ref.~\cite{Gyulassy:1982wignertheory1} pointed out that the ratio between the contribution of intermediate interactions $\left \langle \delta \rho \right \rangle$ and final interaction $\left \langle \rho_{f} \right \rangle$ can be estimated by $\left \langle \delta \rho \right \rangle / \left \langle \rho_{f} \right \rangle \sim (dP)^{-1}(dq_{min})^{-1}$, where $d^{-1}$ is the momentum scale of the Wigner density function of the composite particle, $P$ is the total momentum of the composite particle, and $q_{min}$ is the scale of the relative momentum between particles to characterize the final state kinetic freeze-out. $d^{-1}$ for the $DK$ system is about 1.3 fm, and in our PACIAE simulation,$q_{min}$ is about 1 GeV, and $P$ is about 2.7 GeV, so the ratio  $\left \langle \delta \rho \right \rangle / \left \langle \rho_{f} \right \rangle \sim 0.01$. The three-body system can be treated as a single meson combined with a two-body object, thus we can estimate the ratio for $D\bar{D}K$ in a similar way, which is also about 0.01. The small value indicates that the contribution from intermediate interactions can be neglected.

\subsection{Statistical model}

 In the statistical model~\cite{Becattini:1995if}, the production of hadrons is treated as a rapid phase transition from hot partonic matter to hadronic matter when the chemical equilibrium and thermal equilibrium are reached and then maintained after an expansion of the ``collision fireball". It has been successfully applied to study hadron productions in $e^+e^-$ collisions, in particular, the ground-state charm mesons and charm-strange mesons~\cite{Becattini:1995if}. Here, we extended this framework to calculate the production rates of $D_{s0}^*(2317)$ and $D_{s1}(2460)$ assuming them to be conventional $P$-wave charm strange mesons.  

According to the statistical model~\cite{Becattini:1995if}, the average particle number of a particular hadron in one jet can be calculated from the partition function $Z(\bold{Q})$ of this jet, where $\bold{Q}=(N, S, C, B)$ is the vector for the quantum numbers, i.e., baryon, strangeness, charm, and beauty numbers, of the jet, and their corresponding parameters in the $U(1)$ symmetry group are $\phi=(\phi_1, \phi_2, \phi_3, \phi_4)$. It has the form ~\cite{Becattini:1995if}
\begin{equation}
\begin{aligned}
\langle n \rangle = \frac{1}{Z} \frac{1}{(2 \pi)^4} \int d^4 \phi e^{i \bold{Q} \cdot \phi} \exp{\left \{ \sum_{j=1}^{N_B} \sum_{k} \log(1-e^{-\beta\cdot p_k- i \bold{q}_j \cdot \phi})^{-1} + \sum_{j=1}^{N_F} \sum_{k} \log(1+e^{-\beta\cdot p_k- i \bold{q}_j \cdot \phi}) \right \}} \sum_{k} \frac{1}{e^{\beta \cdot p_k + i \bold{q}_i \cdot \phi} \pm 1}.
\end{aligned}
\end{equation}
where $k$ is all available states in phase space for particle $j$, $p_k$ is the four-momentum of the $k$ state, $\beta$ is the inverse temperature four-vector, $\bold{q}_j=(N_j, S_j, C_j, B_j)$ is the quantum number vector for hadron $j$, $N_B$ and $N_F$ is the number of bosons and fermions, sign ``$-$" in the last term is for bosons and ``$+$" is for fermions.
For continuous level densities, the summation for each hadron can be written as an integral~\cite{Becattini:1995if}
\begin{equation}
\begin{aligned}
\sum_{k} \to (2J+1) \frac{V}{(2\pi)3} \int d^3p,
\end{aligned}
\end{equation}
where $J$ is the spin of the particle. In the center of mass system  $\beta=\left ( 1/T,0,0,0 \right )$. If the temperature of the system is around the energy scale of soft QCD, i.e., $T \backsim 100$ MeV, then all the terms in the exponential factor is much smaller than one except for the term for the pion. Thus the exponential function and natural logarithm function in the integral can be simplified as follows ~\cite{Becattini:1995if}
\begin{equation}
\begin{aligned}
\log \left (1 \pm e^{- \sqrt{p^2+m_i^2}/T - i \bold{q}_i \cdot \phi} \right )^{\pm1}\backsimeq e^{- \sqrt{p^2+m_i^2}/T - i \bold{q}_i \cdot \phi},\\
\frac{1}{e^{- \sqrt{p^2+m_i^2}/T - i \bold{q}_i \cdot \phi} \pm 1 } \backsimeq e^{- \sqrt{p^2+m_i^2}/T - i \bold{q}_i \cdot \phi}.
\end{aligned}
\end{equation}

Now the partition function has the form ~\cite{Becattini:1995if}

\begin{equation}
\begin{aligned}
Z(\bold{Q})=\frac{F_{\pi}}{(2 \pi)^4}  d^4 \phi e^{i \bold{Q} \cdot \phi} \exp{\left \{ \sum_{i} z_i e^{-i \bold{q}_i \cdot \phi} \right\}},
\end{aligned}
\end{equation}

\begin{equation}
\begin{aligned}
z_i=(2J_i+1) \frac{V}{(2 \pi)^3} \int d^3 p e^{- \sqrt{p^2+m_i^2}/T} = (2 J_i+1) \frac{V}{(2 \pi)^3} m_i^2 K_2 (\frac{m_i}{T}),\\
F_{\pi}=\exp{ \left \{ -\sum_{i=1}^3 \frac{V}{(2 \pi)^3} \int d^3 p \log{(1-e^{- \sqrt{p^2+m_i^2}/T})} \right \} },
\end{aligned}
\end{equation} 
where the function $K_2$ is the modified Bessel's function of order two. When the mass of the particle is much larger than the temperature, i.e. $m \gg T$, $K_2$ can be approximated by 
\begin{equation}
\begin{aligned}
K_2 \left( \frac{m}{T} \right) \approx \sqrt{ \frac{\pi T}{2 m} } e^{-m/T}.
\end{aligned}
\end{equation}
Finally the particle number of hadron $i$ in one jet can be furhter simplified as ~\cite{Becattini:1995if}
\begin{equation}
\begin{aligned}
\langle n_i \rangle = z_i \frac{Z(\bold{Q}-\bold{q}_i)}{Z(\bold{Q})},
\end{aligned}
\end{equation}
except for the pion, which reads ~\cite{Becattini:1995if}
\begin{equation}
\begin{aligned}
\langle n_i \rangle = \frac{V}{(2 \pi)^3} \int d^3p \frac{1}{e^{- \sqrt{p^2+m_i^2}/T} -1 }.
\end{aligned}
\end{equation}
For those hadrons containing a $c$ or $b$ quark,  $z_i \ll 1$. Then one can perform an  expansion, $\exp{\{z_i e^{-i\bold{q}_i \cdot \phi }\}}\approx 1+z_i e^{-i\bold{q}_i \cdot \phi }$, and the partition function can be simplified as ~\cite{Becattini:1995if}
\begin{equation}
\begin{aligned}
&Z(\bold{Q})\approx \frac{F_{\pi}}{(2 \pi)^2} \int  d^2 \phi e^{i \bold{Q} \cdot \phi} e^{f(\phi)} \delta_{C,0} \delta_{B,0}+\sum_{i_c} z_{i_c} \frac{F_{\pi}}{(2 \pi)^2}\int  d^2 \phi e^{i (\bold{Q}-\bold{q}_{i_c}) \cdot \phi} e^{f(\phi)} \delta_{C,C_{i_c}} \delta_{B,0}\\
&+\sum_{i_b} z_{i_b} \frac{F_{\pi}}{(2 \pi)^2}\int  d^2 \phi e^{i (\bold{Q}-\bold{q}_{i_b}) \cdot \phi} e^{f(\phi)} \delta_{C,0} \delta_{B,B_{i_b}}+\sum_{i_c,i_b} z_{i_c} z_{i_b} \frac{F_{\pi}}{(2 \pi)^2}\int  d^2 \phi e^{i (\bold{Q}-\bold{q}_{i_c}-\bold{q}_{i_b}) \cdot \phi} e^{f(\phi)} \delta_{C,C_{i_c}} \delta_{B,B_{i_b}},
\end{aligned}
\end{equation}
and
\begin{equation}
\begin{aligned}
f(\phi)=\sum_i z_i e^{-i \bold{q}_i \cdot \phi},
\end{aligned}
\end{equation}
where vector $\bold{Q}=(N, S)$ and $\bold{q}_i=(N_i, S_i)$, charm $C$ and beauty $B$ are shown in Kronecker delta, $i_c$ and $i_b$ are the index for all charm hadrons and bottom hadrons, and $i$ is the index for all the other light hadrons.
\subsubsection{Production of charm mesons in the $e^+e^- \to c\bar{c}$  process}
Considering the production yield of a hadron $i$ in the jet from a $c$ quark in the $e^+e^- \to c\bar{c}$ process, one has~\cite{Becattini:1995if}

\begin{equation}
\begin{aligned}
\langle n_i \rangle = z_i \frac{\int  d^2 \phi e^{i (\bold{Q}-\bold{q}_{i}) \cdot \phi} e^{f(\phi)} }{\sum_{i_c} z_{i_c} \int  d^2 \phi e^{i (\bold{Q}-\bold{q}_{i_c}) \cdot \phi} e^{f(\phi)} }.
\end{aligned}
\end{equation}
We note that the integrals in the numerator and denominator are the same for those hadrons with the same quark constituents. Therefore, for two $D$ mesons containing the same quark constituents but with different quantum numbers,  the ratio of their production yields is 
\begin{equation}
\begin{aligned}
\frac{\langle n_1 \rangle }{\langle n_2 \rangle } = \frac{z_1 }{z_2}.
\end{aligned}
\end{equation}
Because the masses of charm mesons are much larger than the temperature $T$,  the calculation can be much simplified in the following way
\begin{equation}
\begin{aligned}
\frac{\langle n_1 \rangle }{\langle n_2 \rangle }
&=\frac{(2 J_1+1) \frac{VT}{(2 \pi)^3} m_1^2 K_2 (\frac{m_1}{T}) }{(2 J_2+1) \frac{VT}{(2 \pi)^3} m_2^2 K_2 (\frac{m_2}{T}) }\\
&=\frac{(2J_1+1) \frac{VT}{(2 \pi)^3} m_1^2 \sqrt{ \frac{\pi T}{2 m_1} } e^{-m_1/T } }{(2 J_2+1) \frac{VT}{(2  \pi)^3} m_2^2 \sqrt{ \frac{\pi T}{2 m_2} } e^{-m_2/T} }\\
&=\frac{(2 J_1+1) m_1^{\frac{3}{2}} e^{-m_1/T}}{(2 J_2+1) m_2^{\frac{3}{2}} e^{-m_2/T} }.
\end{aligned}
\end{equation}
As shown in this formula, the production ratio of charm mesons is only determined by the spin and mass of hadrons involved, and the temperature of the hadron gas after hadronization. In Ref.~\cite{Becattini:2008tx}, the temperatures of the hadron gas in different processes and energies have been given, and we note that the temperatures in electron-positron annihilations are almost independent of energies and are around a universal value $160-170$ MeV. Thus we use $160$ MeV as the temperature of the hadron gas in our present work.

\subsubsection{Production of $D$ and $D_s$ mesons}
We first compare the production yields of $D$ and $D^*$ mesons obtained from the statistical model with the experimental data and simulation results of PACIAE. In the statistical model, we take the $D$ meson masses from RPP~\cite{Workman:2022ynf}, i.e.,  $m_D=1867.25$ MeV, $m_{D^*}=2008.55$ MeV. The ratio between the production yield of $D^*$ and that of $D$ turns out to be $1.22$. The simulation of PACIAE is performed in the $e^+e^- \to c\bar{c}$ mode and at the center of mass energy of 10.58 GeV. As shown in Table~\ref{tab:dmesons1}, the results from the statistical model and the MC simulation are all consistent with the experimental data at the level of 10\%, which supports the $q\bar{q}$ nature of $D$ and $D^*$.
\begin{table}[htpb]
 \caption{\label{tab:dmesons1}Ratio of the production yields of $D^*$ and $D$ mesons.}
 \setlength{\tabcolsep}{3.2pt}
\begin{tabular}{cc}
 \hline
 \hline
& $f(c\to D^*)/f(c \to D)$\\ \hline
Statistical model&1.22\\
PACIAE simulation&1.18\\
Experimental data~\cite{Lisovyi:2015uqa}&1.28\\
\hline
 \hline
\end{tabular}
\end{table}

Same as the $D$ mesons,  the production yields of  $D_s$ mesons with different quantum numbers can be obtained (For those particles containing an $s$ quark, one should multiply a suppression factor $\gamma_s$ to take into account a non-complete strange chemical equilibrium. But in the ratio, the effect of this factor is canceled). To compare with the experimental data of $D_{s0}^*(2317)$ and $D_{s1}(2460)$ which are obtained for momenta larger than 3.2 GeV/c, we use the MC method (PACIAE) to estimate the percentage $P(p^*>3.2 \ \rm {GeV})$ of  $D_s$ mesons in that momentum range, which is about $40 \%$. From the experimental data, $f(c \to D_s^+)=0.0691$~\cite{Lisovyi:2015uqa}, the production yields of $D_{s0}^*(2317)$, $D_{s1}(2460)$ (their production yields have been multiplied by the percentage $P(p^*>3.2 \ \rm {GeV})$) and $D_{s1}(2536)$ are shown in Table ~\ref{tab:dmesons2} of the main text.  The results show that the production yield of $D_{s1}(2536)$ is close to the experimental value, which implies that the production yields of the $D_s$ mesons are reliable at the level of 30\% if they are genuine $c\bar{s}$ states.
\end{widetext}
\end{document}